\documentclass[sigconf, nonacm]{acmart}
\usepackage{xcolor}
\usepackage{colortbl}
\usepackage{xspace}
\usepackage{pagecolor}
\usepackage{subcaption}
\usepackage{amsmath}
\usepackage{listings}
\usepackage{algorithm}
\usepackage{algpseudocodex}
\usepackage{multirow}
\usepackage{multicol}
\usepackage{tikz}

\usepackage{bbm}
\usepackage{pifont}
\usepackage{hhline}
\usepackage{arydshln}
\usepackage{graphicx}
\usepackage[clock, misc]{ifsym}
\usepackage[inline]{enumitem}
\usepackage[percent]{overpic}
\usepackage{tcolorbox}
\tcbuselibrary{breakable}
\usetikzlibrary{positioning, shapes.geometric, arrows}
\usepackage{adjustbox}
\usepackage{tikz}
\usepackage[normalem]{ulem}
\let\originalinput\input 
\usepackage{soul}

\renewcommand{\input}[1]{%
  \originalinput{#1}
}

\definecolor{darkgreen}{RGB}{0, 100, 0}
\definecolor{experimentblue}{RGB}{127, 159, 186}
\definecolor{experimentgreen}{RGB}{127, 186, 130}
\definecolor{experimentgray}{RGB}{211,211,211}

\definecolor{darkorange}{RGB}{240,131,15}

\newcommand{\baseline}[1]{\textsf{#1}}

\newcommand{\oursystem}{\baseline{LIMAO}\xspace}

\newcommand{\balsaimpl}{\baseline{LIMAO-Balsa}\xspace}

\newcommand{\baoimpl}{\baseline{LIMAO-Bao}\xspace}

\newcommand{\ml}{ML}

\newcommand{\rl}{RL}
\newcommand{\lqo}{LQO}
\newcommand{\lcp}{LCP}
\newcommand{\neo}{Neo}
\newcommand{\bao}{Bao}

\newcommand{\balsa}{Balsa}

\newcommand\vldbdoi{XX.XX/XXX.XX}

\newcommand\vldbvolume{14}
\newcommand\vldbissue{1}

\newcommand\vldbavailabilityurl{URL_TO_YOUR_ARTIFACTS}
\newcommand\vldbpagestyle{plain}

\DeclareRobustCommand{\decomposer}{Plan Decomposer}

\DeclareRobustCommand{\selector}{Task Modules Selector}
\DeclareRobustCommand{\predictor}{Modular Plan Cost Predictor}

\DeclareRobustCommand{\expmanager}{Experience Manager}

\DeclareRobustCommand{\imdbone}{IMDB\_set1}
\DeclareRobustCommand{\imdbtwo}{IMDB\_set2}
\DeclareRobustCommand{\imdbthree}{IMDB\_set3}
\DeclareRobustCommand{\imdbfour}{IMDB\_set4}
\DeclareRobustCommand{\imdbfive}{IMDB\_set5}
\DeclareRobustCommand{\imdbsix}{IMDB\_set6}
\DeclareRobustCommand{\tpchone}{TPC-H\_set1}
\DeclareRobustCommand{\tpchtwo}{TPC-H\_set2}
\DeclareRobustCommand{\tpchthree}{TPC-H\_set3}
\newcommand{\tpch}{TPC-H}

\DeclareRobustCommand{\imdbcombined}{IMDB\_Combined}
\DeclareRobustCommand{\tpchcombined}{TPC-H\_Combined}

\DeclareRobustCommand{\imdbcombined}{IMDB\_Combined}
\DeclareRobustCommand{\tpchcombined}{TPC-H\_Combined}

\DeclareRobustCommand{\truemodel}{$\mathcal{M}$}
\DeclareRobustCommand{\modelcopy}{$\mathcal{M}'$}

\begin{document}
\title{{\oursystem}: A Framework for Lifelong Modular Learned Query Optimization}
\author{Qihan Zhang}
\email{qihanzha@usc.edu}
\affiliation{%
  \institution{University of Southern California}
  \streetaddress{P.O. Box 1212}
  \city{Los Angeles}
  \state{California}
  \country{USA}
}
\author{Shaolin Xie}
\email{shaolinx@usc.edu}
\affiliation{%
  \institution{University of Southern California}
  \streetaddress{P.O. Box 1212}
  \city{Los Angeles}
  \state{California}
  \country{USA}
}
\author{Ibrahim Sabek}
\email{sabek@usc.edu}
\affiliation{%
  \institution{University of Southern California}
  \streetaddress{P.O. Box 1212}
  \city{Los Angeles}
  \state{California}
  \country{USA}
}

\renewcommand{\shortauthors}{Anon. Authors}

\begin{abstract}

Query optimizers are crucial for the performance of database systems. Recently, many learned query optimizers ({\lqo}s) have demonstrated significant performance improvements over traditional optimizers. However, most of them operate under a limited assumption: a static query environment. This limitation prevents them from effectively handling complex, dynamic query environments in real-world scenarios. Extensive retraining can lead to the well-known catastrophic forgetting problem which reduces the {\lqo} generalizability over time. In this paper, we address this limitation and introduce {\oursystem} (\underline{Li}felong \underline{M}odul\underline{a}r Learned Query \underline{O}ptimizer), a framework for lifelong learning of plan cost prediction that can be seamlessly integrated into existing LQOs. {\oursystem} leverages a modular lifelong learning technique, an attention-based neural network composition architecture, and an efficient training paradigm designed to retain prior knowledge while continuously adapting to new environments. We implement {\oursystem} in two {\lqo}s, showing that our approach is agnostic to underlying engines. Experimental results show that \oursystem significantly enhances the performance of LQOs, achieving up to a 40\% improvement in query execution time and reducing the variance of execution time by up to 60\% under dynamic workloads. By leveraging a precise and self-consistent design, \oursystem effectively mitigates catastrophic forgetting, ensuring stable and reliable plan quality over time. Compared to Postgres, \oursystem achieves up to a 4× speedup on selected benchmarks, highlighting its practical advantages in real-world query optimization.

\end{abstract}
\maketitle
\pagestyle{\vldbpagestyle}

\begingroup
\renewcommand\thefootnote{}\footnote{\noindent
This work is licensed under the Creative Commons BY-NC-ND 4.0 International License. Visit \url{https://creativecommons.org/licenses/by-nc-nd/4.0/} to view a copy of this license. For any use beyond those covered by this license, obtain permission by emailing \href{mailto:info@vldb.org}{info@vldb.org}. Copyright is held by the owner/author(s). Publication rights licensed to the VLDB Endowment. \\
\raggedright Proceedings of the VLDB Endowment, Vol. \vldbvolume, No. \vldbissue\ %
ISSN 2150-8097. \\
\href{https://doi.org/\vldbdoi}{doi:\vldbdoi} \\
}\addtocounter{footnote}{-1}\endgroup
\ifdefempty{\vldbavailabilityurl}{}{
\vspace{.3cm}
\begingroup\small\noindent\raggedright\textbf{PVLDB Artifact Availability:}\\
The source code, data, and/or other artifacts have been made available at \url{https://github.com/Tsihan/LIMAOLifeLongRLDB}.
\endgroup
}
\section{Introduction}
\label{sec:introduction}

Query optimizers are crucial components of database systems, responsible for selecting efficient execution plans for queries. Traditional query optimizers have been the backbone of database systems for decades (e.g.,~\cite{selinger1979access, volcano}). However, their inherent limitations and challenges, such as reliance on heuristics (e.g., attribute independence~\cite{lightweightqo, leis2015good}) and inaccurate cost models especially with complex workloads and diverse data distributions, have spurred the development of learned query optimizers ({\lqo}s) (e.g.,~\cite{neo, bao, balsa,kepler, zhu2023lero}), where machine learning ({\ml}) models have been used to replace or improve different components of the query optimizer including cardinality estimation~\cite{learnedce21,deep_card_est2,neurocard}, cost prediction~\cite{learned_cost_estimator_1,learned_cost_estimator_2}, and plan search~\cite{balsa, neo, bao, zhu2023lero}.

While effective in certain scenarios, existing {\lqo}s struggle in dynamic environments with shifting data distributions and query workloads. Recent efforts (e.g.,\cite{robustlero,negi2023robust}) have primarily focused on managing controlled and infrequent shifts. However, addressing significant and frequent changes remains a challenge, as the predominant solution still relies on \textit{complete retraining from scratch}. This approach is computationally expensive and susceptible to performance degradation due to \textit{catastrophic forgetting}\cite{catastrophic}, where previously learned knowledge is lost when adapting to new workloads, particularly in cases of temporary changes or workload reversions.

In this paper, we address this limitation by introducing {\oursystem}, a framework for \underline{Li}felong \underline{M}odul\underline{a}r Learned Query
\underline{O}ptimization. {\oursystem} focuses on the \textit{learned cost prediction} ({\lcp}) problem, where {\ml} models (e.g., tree convolution networks~\cite{tcn} and transformers~\cite{mikhaylov2022learned}) are employed to predict (sub-)plan costs. In particular, {\oursystem} aims at promoting existing {\lcp}s (e.g.,~\cite{balsa,neo,bao,zhu2023lero}) in {\lqo}s to be \textit{lifelong} learners; capable of continuously learning and adapting over time while retaining and leveraging previously acquired knowledge to ensure stable performance. {\oursystem} is based on the idea of learning reusable knowledge in a modular manner~\cite{mendez2022modular,modularblog}. The core intuition is that, to develop a \textit{versatile} {\lcp} capable of accurately predicting the costs of unseen queries, it is essential to learn reusable knowledge of queries’ sub-plans, i.e., tasks, that can be recombined for new queries, rather than tailoring {\lcp}s to specific query types or workloads. This approach draws inspiration from advancements in robotics and mirrors human problem solving, where accumulated, reused, and recombined skills enable efficient handling of novel challenges over time.

Realizing the idea of learned reusable knowledge in the query optimization context requires addressing three key challenges: (1)~identifying a modular decomposition approach suited to the tree structure of query plans, (2)~ensuring the {\lcp} effectively reuses existing knowledge about tasks (i.e., sub-plans) before incorporating new knowledge to prevent redundant learning, and (3)~developing a learning strategy that preserves previously acquired knowledge when integrating new knowledge from tasks in new queries. To address these challenges, {\oursystem} introduces several core components. To tackle the first challenge, {\oursystem} employs an efficient plan decomposer that splits query plans into sub-plans (i.e., tasks) using a new notion of "break" operators that defines modular points in the plan tree. For the second challenge,  {\oursystem} builds and maintains \textit{Module Hubs}, a shared repository of specialized neural network modules corresponding to specific task types, using an efficient variation of the $K$-prototype algorithm~\cite{huang1997clustering}.
{\oursystem} further introduces a new \textit{attention-based neural network composition} architecture that dynamically integrates the selected modules from the hub and assigns appropriate weights to their contributions to the final plan cost estimation.
To overcome the third challenge, {\oursystem} proposes a novel two-phase training approach grounded in the lifelong learning paradigm that avoids overriding previous knowledge. This approach leverages \textit{experience replay} buffers along with a lightweight episodic updates mechanism to facilitate frequent, low-overhead model updates during the online phase while eliminating the need for full retraining in the offline phase. 

{\oursystem} is a generic framework designed to seamlessly integrate with almost all existing {\lqo}s. In this paper, we showcase its potential by integrating it with {\balsa}~\cite{balsa}, a state-of-the-art reinforcement learning-based query optimizer, and \bao, a state-of-the-art hint‑based query optimizer, as system prototypes. We tested on IMDB~\cite{leis2015good} and TPC-H~\cite{tpch} databases using Postgres and designed challenging dynamic situations for them. Our experimental evaluation across various dynamic workloads and data distribution shifts demonstrates that {\oursystem} not only shows better results on \textbf{execution time} but also outperforms the prototype by orders of magnitude in \textbf{query stability} (variance of plan quality over the same query). For IMDB queries, \oursystem improves plan execution time up to $40\%$ and query stability by $60\%$, effectively addressing catastrophic forgetting. \oursystem also achieves up to $400\%$ gains in execution time compared to Postgres. For TPC-H queries, $2\times$ speedup is achieved, and the query stability gain is more than $100\times$, demonstrating significant performance improvements and resilience under dynamic situations. \oursystem can reduce the number of bad plans to 0, while \balsa\ has a few hundred. We also test \oursystem's resilience to overcome the fluctuation in different dynamic levels on different workloads for IMDB, the result shows that \oursystem can achieve up to $6\times$ speedup than \balsa.

\section{{\lcp} as a Lifelong Learning Problem}
\label{sec:problem}

\subsection{{\lcp} in Learned Query Optimizers}
\label{sec:lcp_background}

\begin{figure}
    \centering
    \includegraphics[width=0.9\linewidth]{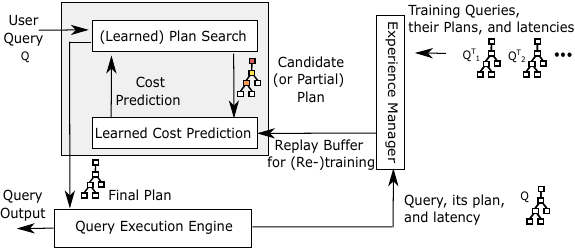}
    \caption{A typical learned query optimizer architecture.}
    \label{fig:lqo_architecture_alone}
    \vspace{-15pt}
\end{figure}

Plan cost prediction is a critical operation in query optimization that determines the efficiency of an execution plan. Traditional cost models rely on formulas that take cardinality estimates of sub-queries as inputs and use manually-tuned constants and heuristics refined over years to align estimated costs with actual performance (e.g.,~\cite{selinger1979access, volcano, cascades}). While effective, these traditional models often struggle to capture the complexity of varying data distributions and query workloads~\cite{learned_cost_estimator_1,learned_cost_estimator_2}. 

Many {\lqo}s (e.g.,~\cite{balsa,neo,bao,zhu2023lero}) adopt \textit{learned cost prediction} ({\lcp}) techniques, where ML models, such as tree convolution networks~\cite{tcn} and attention mechanisms~\cite{attention}, are used to predict the cost of (sub-)plans based on execution statistics from diverse datasets and workloads~\footnote{Unlike approaches that focus solely on learning cardinality estimation~\cite{neurocard,unsupervised_card19,flat,flowloss,querydrivence1,lceindepth,unifiedquerydriven}, our focus is on learned end-to-end cost prediction techniques.}. {\lcp}s play a central role in these {\lqo}s to guide the plan search algorithms to find the best execution plan. For instance, the plan search process can be modeled as a deep reinforcement learning problem, with {\lcp} as value networks to guide searches that complete partial plans into full execution plans~\cite{balsa}. Alternatively, other plan search algorithms learn how to select among candidate plans generated by traditional optimizers while leveraging {\lcp}s to estimate the cost of these candidate plans (e.g.,~\cite{bao}) or their relative rankings (e.g.,~\cite{zhu2023lero}). Figure~\ref{fig:lqo_architecture_alone} shows a typical architecture of an {\lqo} employing {\lcp} with its plan search.

\subsection{{\lcp} with Lifelong Learning}
\label{sec:lcp_lifelong_learning_challenges}

\begin{figure}
  \centering
  \includegraphics[width=0.35\textwidth]{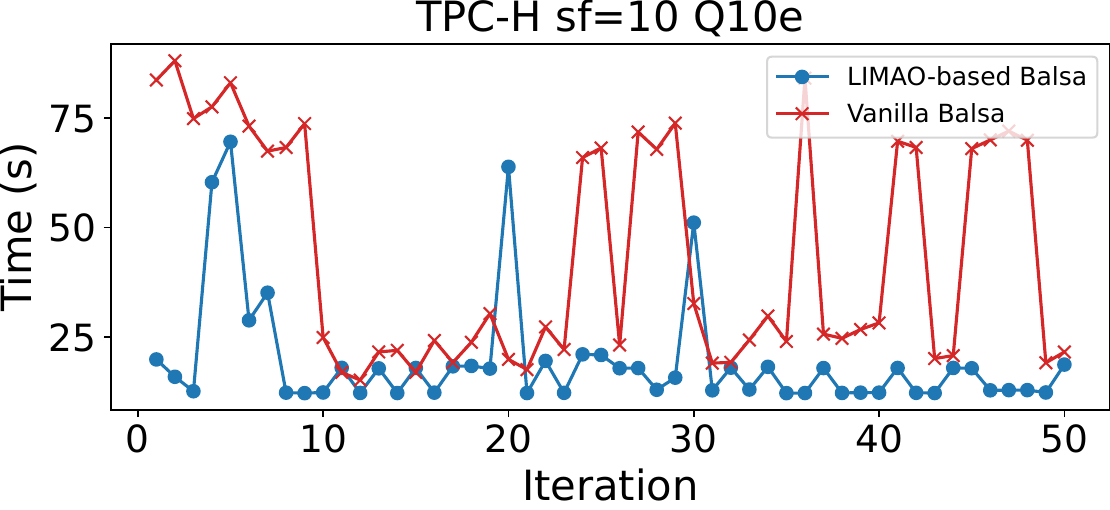}
  \caption{Catastrophic forgetting on the {\lqo}s performance.}
  \label{fig:two_similar_to_lifelongrl_curves}
  \vspace{-5pt}
\end{figure}

\noindent\textbf{Performance Degradation in Dynamic Environments.} While existing {\lcp} approaches (e.g.,~\cite{learned_cost_estimator_1,learned_cost_estimator_2, balsa,neo,bao}) have shown promise, they face significant challenges in dynamic environments of data and query workloads. Some LQOs have attempted to address the dynamic environment challenges by emulating the data and workload drift scenarios during training, such as randomly removing~\cite{robustlero} or partially masking query information~\cite{negi2023robust} to force {\ml} models to make predictions with missing information. However, these approaches are only effective for slight and infrequent changes in query or data distributions. Most {\lcp} approaches still require \textit{complete} (re-)training from scratch (i.e., {\ml} model parameters are completely re-learned) when faced with \textit{significant} and \textit{frequent} changes in the environment. For example, an {\lcp} trained on a workload dominated by JOB queries~\cite{leis2015good} will generate suboptimal plans if the workload is frequently altered with CEB queries~\cite{negi2023robust}, which involve more complex join patterns than JOB. While robust learned cardinality estimators (e.g.,~\cite{negi2023robust,li2023alece}) might adapt to such changes, the {\lcp} still needs to be retrained to reflect this change in the plan cost prediction. Extensive retraining of {\lcp}s in response to frequent changes in dynamic environments can lead to the \textit{catastrophic forgetting} problem~\cite{catastrophic}, where the learned model fails to retain prior knowledge when retrained on new ones, especially in cases of ~\textit{temporary changes} or ~\textit{workload reversions}. 
To better understand this failure behavior, we conducted a simple experiment that simulates frequent workload drift using a {\tpch} query workload~\cite{tpch} (scale factor of 10). We divided the workload into two sets, namely set1 and set2, based on query templates, and alternated their execution every 5 iterations over a total of 100 iterations. In each iteration, all queries from the active set were executed (e.g., set1 ran during iterations 1–5, set2 during iterations 6–10, and so on, with each set running for a total of 50 iterations). We evaluated two variations of {\balsa}~\cite{balsa}, a typical {\rl}-based {\lqo}: vanilla {\balsa}, which suffers catastrophic forgetting, and our {\oursystem}-based {\balsa}, which addresses this issue based on our proposed {\oursystem} framework. Both variations incrementally retrained their {\lcp}s at the end of each iteration. This means that for iterations 2 to 4 within any 5-iteration period, queries used an {\lcp} that had been (re)-trained on their current query set, while in iteration 1, they relied on an {\lcp} trained on the other query set from the previous period. Figure~\ref{fig:two_similar_to_lifelongrl_curves} shows the execution time of one query in set1 across its 50 iterations. After iteration 20, the performance of both variations started to stabilize, where vanilla {\balsa} encounters latency jumps at the beginning of the 5-iterations period due to the catastrophic forgetting of its {\lcp} that was re-trained on the other query set in the previous 5-iterations period. In contrast, the {\oursystem}-based variation was more stable despite frequent workload changes.

\noindent\textbf{Our Goal.} Our ultimate goal is to develop a \textit{lifelong} {\lcp} that can be useful for long stretches of time. The {\lcp} can continuously learn and adapt over time, \textit{integrating new knowledge} from changing queries and data while \textit{retaining and leveraging the previously acquired knowledge} to ensure that performance in prior tasks does not degrade. While single-task learning approaches incorporating techniques like transfer learning~\cite{Ben12} and mechanisms to avoid catastrophic forgetting~\cite{catastrophic2} seem to be a natural fit for our problem, they are inherently task-specific and struggle to generalize to significantly different tasks. 
Multi-task learning (e.g.,~\cite{ella}) offers another direction, where a single model can be learned to share knowledge across diverse types of queries and data scenarios, but this approach is impractical since future queries and data patterns cannot be fully anticipated. Following the same direction, large-scale pre-training on diverse query workloads and data distributions (e.g.,~\cite{zeroshotcostmodel,hilprecht2021one}) can provide a foundation, yet assumes stationary distributions and fails to adapt to the dynamic nature of real-world database environments. Furthermore, pre-trained models are inherently imperfect and require continual feedback from the execution environment to correct unexpected behaviors and maintain effectiveness over time.

\section{\oursystem Overview}
\label{section:arch}

\begin{figure*}[h]
  \centering
  \includegraphics[width=0.9\textwidth ]{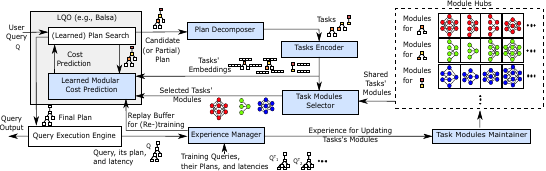}
  \caption{{\oursystem} framework overview.}
  \label{fig:modular_architecture}
\end{figure*}

\noindent\textbf{Main Observation and Challenges.} We observe a key limitation hindering existing {\lcp}s from being suitable for a lifelong setup: they treat each query as a standalone task, which is inadequate for knowledge retention. Queries often comprise fine-grained patterns that are hard to learn with a single {\ml} model. Moreover, these patterns can be shared across queries to help in predicting their costs. For example, in a multi-table join query, it is typical to combine different types of join operators (e.g., nested-loop join, hash join, sort-merge join) to run the query, and such combinations often recur in other queries. Thus, it is valuable to learn these patterns as independent {\ml} \textit{modules}. Learning compositional modules enables the recombination of these modules for unseen queries and the addition of new patterns (i.e., new modules) without requiring a complete update of the {\lcp}'s model. However, achieving this presents three main challenges. First, we need to identify a modular decomposition approach that is suitable for the plan tree structure (\hypertarget{challenge1}{\textbf{Challenge 1}}). Second, we need a strategy that \textit{ensures} that the {\lcp} really reuses knowledge from existing modules to predict the cost of new queries before incorporating any specific new knowledge to these new queries (\hypertarget{challenge2}{\textbf{Challenge 2}}). Third, when incorporating such new knowledge, we must devise a learning strategy that simultaneously prevents the forgetting of knowledge already stored in existing modules (\hypertarget{challenge3}{\textbf{Challenge 3}}).

\noindent\textbf{Framework Overview.} Here, we propose {\oursystem}, a framework (shown in Figure~\ref{fig:modular_architecture}) that tackles the above-mentioned challenges and allows existing {\lcp}s to be efficient lifelong learners. Inspired by recent advances in robotics and human learning processes~\cite{mendez2022modular, modularblog,embodiedmodular,modularhowtoreuse} that focus on learning reusable knowledge, {\oursystem} formulates the learning process of {\lcp} as a \textit{functional compositional} learning problem~\cite{mendez2022modular}. The core idea is to decompose query plans into smaller “tasks” (i.e., sub-plans), each handled by specialized neural network “modules” (e.g., Tree-CNN~\cite{tcn}, Transformer~\cite{attention}), which are then composed into a higher-level network to predict the overall cost (similar to programming, where functions are combined to solve complex problems). For example, a query plan with nested-loop joins and hash joins can reuse pre-trained modules for these operators, recombining them to handle unseen queries efficiently. 

To address~\hyperlink{challenge1}{Challenge 1}, {\oursystem} introduces a {\decomposer} (Section~\ref{sec:plan_decomposition}) that divides the plan into sub-plans (i.e., tasks) based on a new notion of "break" operators. These operators efficiently define points in the plan tree where modular composition can occur. To tackle~\hyperlink{challenge2}{Challenge 2}, {\oursystem} maintains a shared compositional knowledge base called \textit{Module Hubs}. Each hub contains a set of specialized neural network modules that correspond to a specific task type, allowing the model to dynamically select the most relevant modules for a given task. {\oursystem} further employs a composable neural network (Section~\ref{sec:modular_neural_network}) designed to predict the cost of any plan based on the modules chosen for the tasks inside it, dynamically weighting their contributions to refine the final estimate. To overcome~\hyperlink{challenge3}{Challenge 3}, and inspired by prior works~\cite{mendez2022modular}, {\oursystem} adopts an efficient two-phase training paradigm (Section~\ref{sec:q_training}) that encourages knowledge reuse rather than redundantly learning patterns. The training process of {\lcp} is divided into \textit{online} and \textit{offline} phases. During the \textit{online} phase, {\oursystem} attempts to form a cost prediction model for the {\lcp} and train it using only existing \textit{close enough} modules from the Module hubs. To ensure that {\lcp} wisely reuses previously acquired knowledge \textit{without overriding it}, {\oursystem} modifies only a copy of the selected modules using a lightweight training mechanism. In this approach, each training iteration is divided into smaller sections called \textit{episodes}, which allow for prompt model updates and avoid disastrous plans. The original module parameters remain unchanged, and any newly discovered knowledge is stored in a buffer for later use in the offline phase. During the \textit{offline} phase, the {\lcp} incorporates any new knowledge from the buffer by updating the original parameters of existing modules.

Figure \ref{fig:modular_architecture} shows the components and workflow of \oursystem. First, a candidate or a final plan is obtained from an {\lqo}. For example, an {\lqo} like {\bao}~\cite{bao} would rely on complete candidate execution plans generated by a traditional optimizer to select from, whereas {\neo}~\cite{neo} and {\balsa}~\cite{balsa} use a step-by-step plan searcher that builds on partial plans at each iteration. Then, the \textit{\decomposer}\xspace breaks the plan into sub-plans (i.e., tasks) (Section~\ref{sec:plan_decomposition}). Next, the \textit{Tasks Encoder} obtains efficient embeddings for each task using plan-level and operator-level features (Section~\ref{sec:encoding}), which are used by the \textit{\selector}\xspace and \textit{\predictor}\xspace components. The {\selector} (Section~\ref{sec:tasks_modules_selection}) retrieves the most relevant modules from the \textit{Module Hubs} based on embeddings, adopting a $K$-prototype clustering algorithm~\cite{huang1997clustering}. These hubs are created and frequently updated by the \textit{Task Modules Maintainer} (Section~\ref{sec:tasks_modules_selection}). The \textit{{\predictor}} assembles the selected modules into a compositional neural network. Their output query representations are weighted and merged using the attention mechanism~\cite{guo2022attention,de2022attention,brauwers2021general} to generate a final cost estimate for the plan. When the LQO provides a final plan based on the cost estimates, the plan is executed in the query execution engine and feedback is sent to \textit{{\expmanager}} for (re-)training the \textit{\predictor}\xspace and the selected modules using our two-phase training approach. 

\vspace{-5pt}
\section{Plan Decomposition into Tasks}
\label{sec:plan_decomposition}

In this section, we illustrate the process of decomposing a query plan into a set of tasks to address~\hyperlink{challenge1}{\textbf{Challenge 1}}. This process is an important step in enabling the functional composition of {\oursystem}. However, it is very challenging because decomposing a plan into a set of combinable tasks is inherently complex. This complexity arises from the variable structure of the query plans, where the number, types, sequence of operators, and tree depth are not fixed.

\noindent\textbf{Approach.} A straightforward idea for decomposing a query plan is to divide it into pipelines based on "blocking" operators~\cite{meng2007approach}. However, this often results in suboptimal tasks from the functional composition perspective, as blocking operators may create pipelines that are too coarse-grained, resulting in large, monolithic tasks where diverse execution patterns (e.g., various join types or scan methods) are grouped together. Such coarse decomposition makes it challenging to construct reusable modules for other queries. Determining appropriate \textit{breakpoints} in the plan tree for task decomposition is, in general, a non-trivial problem. Sub-trees that are too deep result in ineffective decomposition, as they closely resemble the original query plan, while overly frequent breakpoints create shallow sub-trees that fail to capture plan complexity and add computational overhead due to the proliferation of sub-trees.

To address this, we introduce the concept of "break" operators to guide the plan decomposition process. A break operator is an operator that splits the plan tree at its first occurrence during a top-down traversal. Specifically, we identify the first occurrence of a break operator in both the \textit{left} and \textit{right} sub-trees of the root, and extract the corresponding sub-trees \textit{rooted} at these operators. We stop the traversal upon encountering the first instance of each break operator type to avoid creating structurally overlapping sub-trees rooted at the same operator type. Continuing the decomposition beyond the first occurrence would lead to multiple sub-trees that share the same break operator type at their roots—i.e., they would be \textit{functionally overlapping}—resulting in unnecessary tasks.

\noindent{\textbf{Joins as Break Operators.} In principle, break operators can be any type of query plan operator (e.g., scan, join, aggregate, sort). However, for break operators to be effective, they should satisfy two key criteria: 1) \textit{Performance-critical}: the break operator should significantly influence the query execution time, and 2) \textit{Decomposition-effective}: splitting around this operator should produce sub-plans that are neither too fine-grained nor too coarse, enabling meaningful modular learning. We chose the join operators as break operators because they best satisfy both criteria. First, join operators are typically the most computationally expensive components in query plans.
Second, decomposing around join operators results in sub-plans that are rich in structure and execution variability.}

 \begin{figure}
  \centering
  \includegraphics[width=0.4\textwidth]{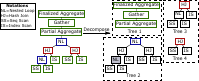}
  \caption{Example on plan decomposition into tasks. Grey operators are those break operators that are ignored because their ancestral nodes already include at least one instance of the same break operator type.}
  \label{fig:new_treedecompose}
  \vspace{-5pt}
\end{figure}

\noindent\textbf{Example.} Figure~\ref{fig:new_treedecompose} shows an example of our plan decomposition approach using two example break operators: hash join (HJ) and nested-loop join (NL). The left side shows the original plan tree, while the right side shows the resulting four sub-trees (i.e., tasks): three sub-trees that are generated by the break operators and one sub-tree that spans the remaining operators (i.e., aggregations, scans, non-break join operators) from the root to the first encountered break operator or leaf nodes for completeness. The plan tree is traversed independently for each break operator type to generate their corresponding tasks. For the NL operator, only one sub-tree (Tree 2) is generated as the first encountered NL node has no sibling and already covers its grandchild NL node, which is colored in grey. Conversely, for the HJ operator, two HJ-rooted sub-trees (Trees 3 and 4) are produced, as the traversal continues through the left and right sub-trees of the NL node, marked in purple. Note that, since Tree 2 is NL-rooted and Tree 3 (or Tree 4) is HJ-rooted (i.e., functionally non-overlapping), they can overlap structurally.
\section{Task Encoding}
\label{sec:encoding}

Efficient encoding of decomposed tasks is crucial to capture their characteristics during the module selection and cost prediction steps. Below, we provide the details of encoding a single task:

\noindent\textbf{Table Selectivity (Feature~$A$).} This captures the selectivity of tables involved in the task. Following the approach in~\cite{negi2023robust}, we use traditional cardinality estimators to compute these selectivities. These estimators add minimal overhead to the inference time of the learned model while providing valuable information. If there are $n$ tables in a workload, then we represent this feature with a numerical vector of length $n$, where each entry contains a normalized value between 0 and 1 (i.e., table selectivity divided by table size) if the task involves the corresponding table, and 0 otherwise.

\noindent\textbf{Operator Mapping (Feature~$B$).} This encodes the join and scan properties associated with the task. To generate it, we perform a bottom-up traversal of the sub-tree corresponding to the task and record the count of each type of join operator (hash join (HJ), merge join (MJ), and nested-loop (NL) join) and scan operator (sequential scan (SS) and index scan (IS)) encountered. If there are $n$ tables in a workload, then we represent this feature with a numerical vector of length $3+2n$, where the first three entries correspond to the counts of the join types, and the remaining entries represent the counts of scan types for each table.

\noindent\textbf{Preorder Index (Feature~$C$).} This encodes the tree structure of the sub-tree associated with the task in a numerical vector. Similar to~\cite{neo,balsa}, we generate this vector using a pre-order traversal of the sub-tree, where each entry represents the index of a node in the original query plan to which the task belongs. For nodes with no children, the corresponding child entries in the vector are set to 0.

\noindent\textbf{Query Feature (Feature~$D$).} This captures whether the SQL query associated with the task possesses any of the following four properties: (1)~contains a sub-query, (2)~includes an aggregation function, (3)~contains a GROUP BY clause, or (4)~contains an ORDER BY clause. It is represented as a binary 0-1 vector of length 4.

\begin{figure}
  \centering
  \includegraphics[width=0.38\textwidth]{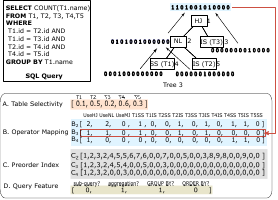}
  \caption{Encoding for the tasks represented with trees~2,~3, and~4 in Figure~\ref{fig:new_treedecompose}.}
  \label{fig:encoding}
\end{figure}

\noindent\textbf{Example.} Figure~\ref{fig:encoding} depicts the encoding of the three break-operator-based tasks in Figure~\ref{fig:new_treedecompose} (we omitted the encoding details of Tree~1 to avoid cluttered visualization). The upper left part shows the SQL query corresponding to the decomposed plan. Features~$A$ and~$D$ are defined at the query-level and then used with all tasks. In contrast, features~$B$ and~$C$ are task-specific, with $B_i$ and $C_i$ representing the features for the task associated with Tree~$i$ ($i \in \{2, 3, 4\}$). Additionally, we detail the process of generating Feature~$B$ for Tree~3, demonstrating a bottom-up traversal that incrementally counts the join and scan operators encountered.

\section{Modules Maintenance and Selection}
\label{sec:tasks_modules_selection}

Identifying the appropriate neural network modules for the decomposed tasks is a critical step in constructing an accurate {\lcp}. A straightforward approach is to treat each unique combination or sequence of operators in a query plan as an independent task and assign a dedicated neural network module to it in the module hub. For example, a task that involves SS followed by NL would require a different module from a task that involves SS followed by "indexed" NL. Similarly, a task executing HJ before NL would require a distinct module from the one executing them in reverse order. However, this approach is impractical, as it would lead to an explosion in the number of modules, significantly increasing storage and search overhead while limiting the benefits of modularization by over-specializing each module to a highly specific task. To address this, our idea is to maintain a \textit{representative} module for each group of \textit{similar enough} tasks using a clustering algorithm. These representative modules serve as a carrier of knowledge, which is the prerequisite for solving~\hyperlink{challenge2}{\textbf{Challenge 2}}. All representative modules built for an {\lqo} integrated with {\oursystem} share the same neural network architecture, inherited from the base LCP model used by this {\lqo} (e.g., the Tree-CNN architecture in Balsa~\cite{balsa}). However, they differ in their parameters, which are specialized through training on their respective tasks (i.e., sub-plans). These parameters are then dynamically adapted as tasks evolve over time, resulting in a balance between accuracy and efficiency.

\noindent\textbf{Building Each Module Hub as a Cluster.} We adopt a variation of the $K$-prototype algorithm~\cite{huang1997clustering} to maintain $K$ representative neural network modules for each type of task decomposed using a specific break operator type (i.e., a set of $K$ representatives for NL-rooted tasks, and another set of $K$ representatives for HJ-rooted tasks, etc). We choose the $K$-prototype algorithm for two key reasons. First, it is an extension of $K$-means designed to handle mixed data types (numerical and categorical features), which is compatible with our numerical and categorical task encodings (Section~\ref{sec:encoding}). Second, it is computationally efficient, making it practical for maintaining and updating module clusters dynamically over time. Our algorithm for constructing clusters for each module hub operates in four steps during the training phase of any {\lqo}. First, initialize $K$ module representatives (i.e., centroids) with random values for Feature~$A$ as the numerical feature, and Features $B$, $C$, $D$ as categorical ones, based on the training set queries. Second, for each task $x_i$ decomposed from the training examples, assign it to the cluster with the smallest dissimilarity value, calculated according to this formula:

\begin{equation}
d = \sum_{j \in \{\text{A}\}} \text{dis}_{\text{num}}(x_{ij}, c_{kj}) + \gamma \sum_{j \in \{\text{B, C, D}\}} \text{dis}_{\text{cat}}(x_{ij}, c_{kj})
\label{eq:cluster_dissimilarity}
\end{equation}

 where $x_{ij}$ is a feature vector (numerical or categorical), and $c_k$ is the representative for cluster $k \in \{1, \dots, K\}$, $\gamma$ is a weighting factor to balance the influence of numerical and categorical features, and $\text{dis}_{\text{num}}$ and $\text{dis}_{\text{cat}}$ are \textit{vector-wise} numerical (e.g., Euclidean distance) and categorical (e.g., mismatch count) dissimilarity functions, respectively. 
Third, for each cluster representative $c_{k}$, update its numerical $c_{kj}^{\text{num}}$ and categorical $c_{kj}^{\text{cat}}$ features using the numerical mean $c_{kj}^{\text{num}} = \frac{1}{|S_k|} \sum_{x_i \in S_k} x_{ij}$ and 
most frequent value $c_{kj}^{\text{cat}} = \arg \max_{v} \sum_{x_i \in S_k} 1(x_{ij} = v)$ functions, respectively, where $S_k$ is the set of tasks assigned to the cluster. Fourth, repeat the second and third steps until convergence; either no change in task assignments occurs or the maximum number of iterations is reached.

\noindent\textbf{Selecting Modules and Updating Hubs.} During the testing phase, we simply use Equation~\ref{eq:cluster_dissimilarity} to determine the appropriate representative cluster for any incoming task that minimizes the dissimilarity value and then retrieve its corresponding neural network module. If the dissimilarity of a task to all clusters exceeds a predefined threshold, the \textit{Task Modules Maintainer} creates a new cluster with that task as its representative. Moreover, the modules maintainer periodically checks cluster sizes and removes any clusters that contain fewer tasks than a specified minimum. 
\section{Modular Plan Cost Prediction}
\label{sec:modular_neural_network}

\begin{figure*}[h]
  \centering
  \includegraphics[width=0.87\textwidth]{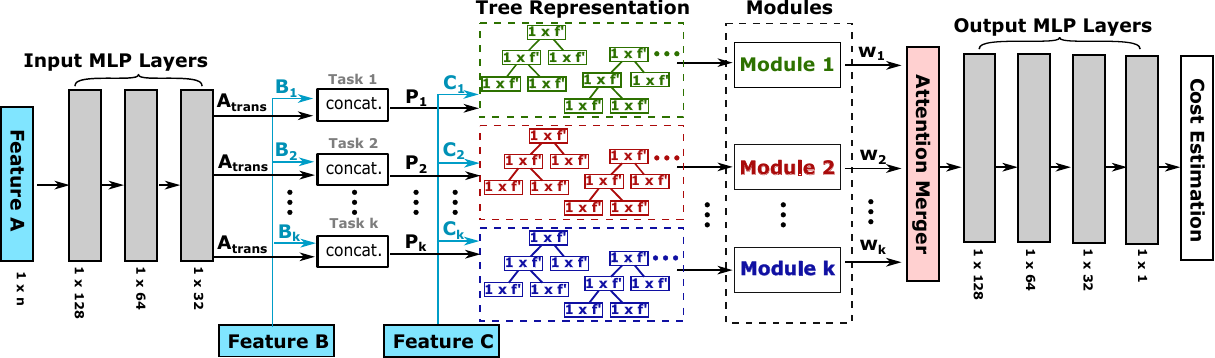}
  \caption{Neural network architecture of modular {\lcp}.}
  \label{fig:attentionassigner}
\end{figure*}

Our primary objective in {\oursystem} is to enable {\lqo}s to construct a customized {\lcp} neural network policy for each query by selecting and combining modules from several shared module hubs, thereby promoting knowledge reuse to address~\hyperlink{challenge2}{\textbf{Challenge 2}}. To achieve this, we propose to use \textit{neural composition} techniques (e.g.,\cite{neuralmodulenet,goyal2021recurrent}) to assemble a complete {\lcp} neural network from the selected modules to handle the query. Neural composition has been extensively applied to build modular architectures in supervised learning (e.g.,\cite{neuralmodulenet,routingnet}) and reinforcement learning (e.g.,\cite{goyal2021recurrent,mtlrlsoft}) and has recently been adapted for functional compositional problems~\cite{mendez2022modular,robottransfer}.

\noindent\textbf{Attention-based Neural Composition.} A large class of existing neural composition methods assumes a linear dependency structure between decomposed tasks (i.e., a chaining or graph structure), which is common in robotic tasks~\cite{mendez2022modular,optioncritic} (e.g., a pick-and-place robotic task is decomposed into sequential steps such as "grasp," followed by "lift," and then "place"). This chaining structure, however, creates brittle dependencies among modules, where changes to one module can have cascading effects on others. Moreover, this assumption does not align well with our break-operator-based decomposition, which can produce non-temporally related tasks. For instance, a multi-table join query may be decomposed into two parallel HJ tasks, as shown in Figure~\ref{fig:new_treedecompose}. Although parallel neural composition approaches~\cite{parallel1,parallel2} seem well-suited for such scenarios, they often fail to capture the inherent inter-dependencies between query tasks, particularly as the outputs of these tasks must be combined to produce a unified cost prediction for the entire query plan. Additionally, not all tasks contribute equally to the final cost prediction, which these methods typically overlook.

To address these limitations, we propose an \textit{attention-based} variation of parallel neural composition. This variation facilitates the composition of cost prediction tasks by leveraging attention mechanisms~\cite{guo2022attention,de2022attention,brauwers2021general} to capture task relationships and adjust their contributions accordingly. The core idea is to assign importance to the outputs of different modules using attention weights, dynamically adjusting their contributions based on their potential impact on the final predicted cost. Attention mechanisms have already demonstrated their effectiveness in other database applications~\cite{li2023alece,lsched}, and we extend their utility to our cost prediction framework. Formally, assume that there are $k$ modules with importance $w_1, \cdots, w_k$, then the assigned weight $w_i'$ for $w_i,i\in [1,k]$ is in a softmax weighting form: \( w_i' = \exp(w_i)/\sum_{j=1}^k \exp(w_j)\).

Furthermore, we train the combination of modules corresponding to one query plan as a whole unit. In this way, the modules are not only trained on their tasks but also learn to collaborate with other modules under different combinatorial setups.

\noindent\textbf{Network Architecture.} Figure~\ref{fig:attentionassigner} shows the neural network architecture of our {\lcp}. After a plan is decomposed into $K$ tasks and their corresponding modules are retrieved (see Section~\ref{sec:tasks_modules_selection}), the table selectivity encoding (i.e., Feature $A$) is passed through a series of fully-connected layers, each progressively reducing in size. The output of the third fully connected layer is then concatenated with the operator mapping encoding (Feature $B$) of each decomposed task, resulting in a new vector representation $P$ for each task. Next, using the pre-order index encoding (Feature $C$), the tree representation of encodings is reconstructed for each task. Each node in this tree corresponds to the previously concatenated vectors from Features $A$ and $B$, resulting in $1 \times f'$ dimension. This tree representation is then processed by its corresponding module to generate a module-specific plan representation $w_i$ for each task (e.g., Tree-CNN-based~\cite{marcus2021bao,negi2023robust,zhu2023lero} or Tree-LSTM-based~\cite{chen2023loger, yu2022hybridqo} plan representation). Afterward, the attention merger is applied to these module-specific plan representations. It dynamically assigns weights to each representation and combines them to send to the output MLP layers, producing the final cost estimation of the plan.

\section{{\oursystem} Training}
\label{sec:q_training}

\begin{algorithm}
\scriptsize
\caption{Training Mechanism in \text{\textit{\oursystem}}} 
\begin{algorithmic}[1]
\State $T \gets 0$; $\mathcal{B}_{episode} \gets \text{[]}$; $ \mathcal{B}_{all} \gets \text{[]}$ \Comment{Initialization}

    \State $\mathcal{M} \gets \text{Init}(\text{input \& output MLP, Module Hubs})$

\While{ $T < \text{iterations}$}

        \If{$T > 0$}  
        
\State $\mathcal{M} \gets \text{DeepCopy}(\mathcal{M}')$ \Comment{Offline updates}
        
            \If{$\text{Drift Detected}$} 
                \State $\mathcal{M}.\text{CompletelyTrainWith}(\mathcal{B}_{all})$ 
            \Else
                \State $\mathcal{M}.\text{CompletelyTrainWith}(\mathcal{B}_{last})$ 
            \EndIf
    \EndIf
        \State $\mathcal{M}' \gets \text{DeepCopy}(\mathcal{M})$, $ \mathcal{B}_{last} \gets \text{[]}$

        \State $e_1 \dots e_n  \gets \text{iteration queries} $ \Comment{Split queries into episodes}  
    
        \For {each episode $e_i$} \Comment{Online exploration phase}
            \For {each query $q_j \in e_i$}
                \State $\textit{modulesIds} \gets \text{ModuleSelector}(q_j)$

                \State $\textit{feedback} \gets \text{PredictAndExecute}(\mathcal{M}', q_j, \textit{modulesIds})$
                \State $\mathcal{B}_{episode}.\text{Add}(\textit{feedback}, \textit{modulesIds})$

            \EndFor

            \State $\mathcal{M}'.\text{LightlyTrainWith}(\mathcal{E})$
            \State 
         $\mathcal{B}_{last}.\text{Add}(\mathcal{B}_{episode})$ and \text{Reset }$\mathcal{B}_{episode}$ 
        \EndFor

    \State $ \mathcal{B}_{all}.\text{Add}(\mathcal{B}_{last}) $
    \State $\text{Evaluate}(\mathcal{M}, \textit{testSet})$ \Comment{Evaluation}

    \State $T \gets T + 1$
\EndWhile
\end{algorithmic}
\label{algorithm:newmodularrl}
\vspace{-2pt}
\end{algorithm}

\oursystem introduces a novel training mechanism that follows the lifelong learning paradigm, incorporating experience buffers and episodic updates to improve the quality of LCP and reduce training costs. The training mechanism effectively addresses ~\hyperlink{challenge3}{\textbf{Challenge~3}} by: 1) taking advantage of reusable modules to eliminate the time needed for retraining from scratch when query or data pattern drifts; 2) using experience replay to quickly reclaim previous knowledge; and 3) dividing training queries within each iteration into subsets of episodes and performing rapid lightweight model training after each episode\footnote{Note that, in {\oursystem}, training iterations and their corresponding episodes use different sets of queries, unlike existing {\lqo}s (e.g.,\cite{balsa, marcus2021bao}), which fix the same set of queries across all training iterations.} In such a way, \oursystem generates better plans in shorter learning periods while effectively avoiding bad plans and stabilizing plan quality. We have two main phases: an \textit{online phase} where we episodically update a copy of the LCP model and populate the experience buffer, and an \textit{offline phase} where we update the original LCP model and perform experience replay. The details of our method (Algorithm~\ref{algorithm:newmodularrl}) are as follows:

\noindent\textbf{Initialization.} 
The \textit{\expmanager}\ maintains an episode buffer $\mathcal{B}_{episode}$ that is refreshed per training episode, and an experience buffer $\mathcal{B}_{all}$ that contains experience from every iteration (line 1). The experience from the previous iteration is defined as $\mathcal{B}_{last}$. We create the LCP model \truemodel\ by initializing the neural network's MLP input and output layers with random parameters, as well as all the modules in Module Hubs and Attention Merger (line 2). To avoid catastrophic damage to the model \truemodel\ while (re-)training it, we maintain a copy \modelcopy\ that is updated frequently in our \textit{episodic} setup, whereas \truemodel\ is updated once per iteration (Details are below).

\noindent\textbf{Online Phase.} An under-trained LCP model, particularly in early iterations, could lead to a sequence of inefficient plans due to infrequent model updates. To avoid this, we introduce an episode-based paradigm that updates the LCP model more frequently to stabilize performance. In each episode, the current LCP model is used to predict query plans for the episode's queries, which are then executed. The resulting runtime feedback, along with the corresponding module combinations, is stored in the episode buffer $\mathcal{B}_{\text{episode}}$ (lines 14–16). At the end of the episode, \modelcopy\ is quickly trained based on the buffer, with only a few epochs of neural network updates (line 17). This lightweight training process introduces minimal latency to the online exploration phase, typically completes in a few seconds, while avoiding damage to the model \truemodel.

\noindent\textbf{Offline Phase.} In this phase, we perform comprehensive training on the model \truemodel\ to ensure it remains up-to-date in dynamic environments (lines 5-9). We start by replacing \truemodel\ with the model copy \modelcopy\, which was episodically trained in the previous iteration. Following this, \truemodel\ is retrained using an experience replay buffer, determined by whether a drift is detected or not. If a drift in query patterns or data distribution is identified~\cite{duong2018high,baidari2021bhattacharyya,hu2022cadm,mallick2022matchmaker,abbasi2021elstream,li2022warper}, then \truemodel\ is retrained using the experience buffer $\mathcal{B}_{all}$, which contains experiences from all prior iterations, allowing the model to retain historical knowledge while integrating new patterns. If no drift is detected, then \truemodel\ only requires light adaptation. In this case, we limit the retraining of \truemodel\ to the most recent experiences stored in $\mathcal{B}_{last}$ (lines 6-9) only to reduce overhead and prevent overfitting. Note that, regardless of whether retraining uses $\mathcal{B}_{all}$ or $\mathcal{B}_{last}$, \truemodel\ converges quickly, as it is initialized from \modelcopy, which has already been lightly exposed to recent workload and data shifts.

\section{Experimental Evaluation}
\label{sec:evaluation}

Here, we conduct experiments to answer some key questions: Can \oursystem help generate better plans in dynamic environments (Section~\ref{subsubsection: compare_other_optimizers})? How does \oursystem training paradigm prevent catastrophic forgetting (Section~\ref{subsubsection: challenging situation})?  How do the design components of \oursystem contribute to its performance (Section~\ref{subsubsection:ablation_study})? How effectively can \oursystem avoid bad plans (Section~\ref{section:Disastrous_plans})? 
\vspace{-4pt}

\subsection{\textbf{Experiment Setup}}
\label{exp_setup}

\noindent\textbf{Benchmarks.} We perform our evaluation using TPC-H~\cite{tpch} and IMDB-based~\cite{leis2015good} benchmarks. For IMDB, we evaluate on the JOB~\cite{leis2015good} and CEB~\cite{negi2023robust} query sets,  as well as a set of queries from~\cite{marcus2021bao}, which we refer to as BaoQs. We construct six benchmarks for IMDB and three benchmarks for TPC-H, where we randomly split each benchmark into training and test sets, as shown in Table~\ref{tab:all_workloads}. The first three IMDB workloads consist of distinct query templates, while the last three reuse the same templates as the first three but apply different filtering conditions to create different queries.

\begin{table}[h!]
 \vspace{-4pt}
    \centering
    \small
    \caption{Evaluation workloads and their templates.}
    \begin{tabular}{cccc}
        \toprule
         Workload & training  & test & Template\\ 
        \midrule
         \imdbone & 94 & 18 & Most from JOB, 2 from BaoQs \\
         \imdbtwo & 71 & 14 & Most from BaoQs, 3 from JOB\\
         \imdbthree & 30 & 5 & Most from CEB, 3 from JOB\\
        \imdbfour & 94 & 18 & Most from JOB, 2 from BaoQs \\
         \imdbfive & 14 & 5 & Most from CEB, 3 from JOB\\
         \imdbsix & 14 & 5 & Most from CEB, 3 from JOB\\
         \tpchone & 34 & 6 & TPC-H template 3, 8, 10, 13 \\
         \tpchtwo & 34 & 6 & TPC-H template 5, 
         7, 12, 14 \\
        \tpchthree &  40 &  12 &  Other TPC-H template except 15 \\
        \bottomrule
       
    \end{tabular}
     \vspace{-4pt}
    \label{tab:all_workloads}
\end{table}

\noindent\textbf{Static and Dynamic Environments.}
 We conducted experiments under four different environments: \textbf{\textit{Static}}, \textbf{\textit{Workload Switch}}, \textbf{\textit{Volume Switch}}, and \textbf{\textit{Both Switch}}. For \textit{Workload Switch}, the experiment iteratively switches between multiple query sets. We refer to the collection of these query sets as a \textit{combined workload}.
 For \textit{Volume Switch}, the experiment iteratively switches between two databases with different data volumes. For the IMDB dataset, this involves switching between the full database and a subset containing only data from the year 2000 onward. For TPC-H, we switch between databases with scale factors of 10 and 1. The \textit{Both Switch} scenario combines changes in both query patterns and data volume simultaneously. Unless otherwise specified, dynamic scenarios involve periodic switching every 5 iterations over a total of 100 iterations, resulting in 50 combined iterations to evaluate for each switch side. For example, in \textit{Workload Switch}, {\imdbone} is used for iterations 1–5, {\imdbtwo} for iterations 6–10, then back to {\imdbone} for iterations 11–15, and so on. We analyze the performance of {\oursystem} under non-periodic switching in Section~\ref{subsubsection:non_periodic} and~\ref{subsubsection:non-periodic-bao}.

 We compare \oursystem to the following LQOs and Postgres:

\noindent\textbf{Baselines:} We evaluate the performance of {\oursystem} against three baselines:
 \noindent\textit{\underline{Balsa}~\cite{yang2022balsa}}, an {\lqo} that uses deep reinforcement learning to construct query plans step-by-step from scratch;
 \noindent\textit{\underline{Bao}~\cite{marcus2021bao}}, an {\lqo} that employs a multi-armed bandit model to steer the traditional optimizer toward more efficient plans via ranked hint sets, without fully replacing it; and
 \noindent\textit{\underline{Postgres}~\cite{postgresql}}, the traditional rule-based and cost-based optimizer (Version 12.5). We integrated {\oursystem} with both Balsa and Bao, denoted as \textbf{\balsaimpl} and \textbf{\baoimpl}, respectively. We also extended Balsa’s source code to support non-equality joins and multiple join conditions between two tables, and applied these improvements to both Balsa and {\balsaimpl}. Our experimental evaluation primarily uses {\balsaimpl} to assess {\oursystem}, while {\baoimpl} is used to explore the performance under non-periodic and cross-schema switching situations in Section~\ref{subsubsection:non_periodic} and Section~\ref{subsubsection:non-periodic-bao}, respectively.

\noindent\textbf{Performance Metrics.} 
We evaluate performance using several key metrics, including query \textit{execution time}, the \textit{number of bad query plans}, and \textit{execution stability}. Execution stability is used to quantify the degree of catastrophic forgetting and is measured using two indicators: the \textit{variance} and the \textit{derivative} of the total workload execution time across iterations. Variance captures fluctuations in execution time, while the derivative is computed from a smoothed curve of execution time to reflect the rate of change over time. Better stability implies better handling of catastrophic forgetting.

\noindent\textbf{Default Settings.} Unless otherwise specified, we adopt the following default configurations. To construct the \textit{Module Hubs} in {\oursystem}, we select HJ and NL as break operators for the IMDB benchmark, with module hub sizes (i.e., $K$ representative modules) of 2 and 3 respectively. For the TPC-H benchmark, we use HJ and MJ as break operators, each with a module hub size of 1. The sizes of \textit{Module Hubs} are determined by the complexity of the workload and the occurrence frequency of the corresponding operator types. In both benchmarks, we include an additional module hub, referred to as \textit{OTH}, which stores modules corresponding to sub-plans not rooted at a break operator (e.g., Tree~1 in Figure~\ref{fig:new_treedecompose}). For training {\oursystem}, we divide the queries in each training iteration into episodes of 10 queries. In both the {\bao} and {\baoimpl} implementations, we use 49 hint sets and train using the most recent 500 execution records.

\noindent\textbf{Machine Settings.}  Unless otherwise specified, experiments are conducted on CloudLab~\cite{cloudlab}, using c240g5 which is equipped with dual Intel Xeon Silver 4114 CPUs (2.20 GHz, 20 cores total), 192 GB DDR4 RAM, a 480 GB SATA SSD, and NVIDIA P100 GPU.

\subsection{End-to-End Experimental Results} \label{subsection:plan_evaluations}

In this section, we evaluate the performance of \oursystem across four dynamic scenarios.
 Section~\ref{subsubsection: compare_other_optimizers} presents the performance of \balsaimpl under a basic \textit{Workload Switch} scenario, where two workloads are alternated periodically. In this setting, \balsaimpl performs comparably to Balsa and shows advantages in certain cases only. Therefore, in Section~\ref{subsubsection: challenging situation} and Section~\ref{subsubsection:non_periodic}, we devise more complex \textit{Workload Switch} scenarios which involve periodic switching among three workloads and the other using a non-periodic switch pattern. In Section~\ref{subsubsection:non-periodic-bao}, we shift to {\baoimpl} to evaluate performance under a non-periodic, cross-schema switching scenario involving five different workloads.
Given the stronger performance of \balsaimpl, we further evaluate its behavior in Sections~\ref{subsubsection:performance comparison over iterations on four situations} to provide a deeper analysis of the benefits introduced by integrating \oursystem into Balsa.

\subsubsection{Comparison with Balsa under a simple \textit{Workload Switch}.}\label{subsubsection: compare_other_optimizers}

In a simple \textit{Workload Switch} scenario, we evaluate the performance of \balsaimpl and Balsa by alternating between \imdbone\ and \imdbtwo\ every 5 iterations over a total of 100 iterations. Table~\ref{tab:best_performance} shows the speedup ratios of the best query plans generated by \balsaimpl and Balsa in this setup. Both systems outperform Postgres by up to $2.4\times$ on IMDB queries. The performance gain in IMDB workloads is largely due to the ability of LQOs like Balsa and \balsaimpl, which generate plans from scratch, to better handle complex join conditions commonly found in these queries. In \imdbone, \oursystem further enhances performance by decomposing complex queries into simpler sub-queries and effectively preserving historical knowledge from the workload, giving it an edge over Balsa. In \imdbtwo, the BaoQs workload is more complex than JOB, featuring more join operations and longer average query execution time. Under such conditions, the limited number of training iterations and queries may not provide enough examples per module combination to train a fully effective model in \balsaimpl, resulting in a slight performance degradation in \imdbtwo\ compared to \imdbone.

For TPC-H, the performance advantage of \balsaimpl and Balsa over Postgres is less obvious, particularly due to the relatively simple join conditions. This observation aligns with prior findings~\cite{doshi2023kepler,yang2022balsa,zhu2023lero}, which show that LQOs typically offer limited improvements over traditional query optimizers for the TPC-H benchmark. Despite this, \balsaimpl still achieves improved performance under the \tpchone\ and \tpchcombined\ workloads. The lack of performance gains in \tpchtwo\ for both Balsa and \balsaimpl can be attributed to operator selection behavior. In \tpchone, using only HJ is generally sufficient to generate optimal plans. However, in \tpchtwo\ which includes templates such as 5, 7, 12, and 14, the use of NL would be more optimal. However, when Balsa becomes biased toward always selecting HJ during training, it fails to consider NL, which may be critical for optimal performance in certain queries. This bias can result in performance that is even worse than Postgres. Increasing the number of training iterations could help mitigate this issue by enabling the model to learn when using NL is beneficial.

\begin{figure}
\hspace*{-1.2cm}
  \begin{center}
  \includegraphics[width=3in]{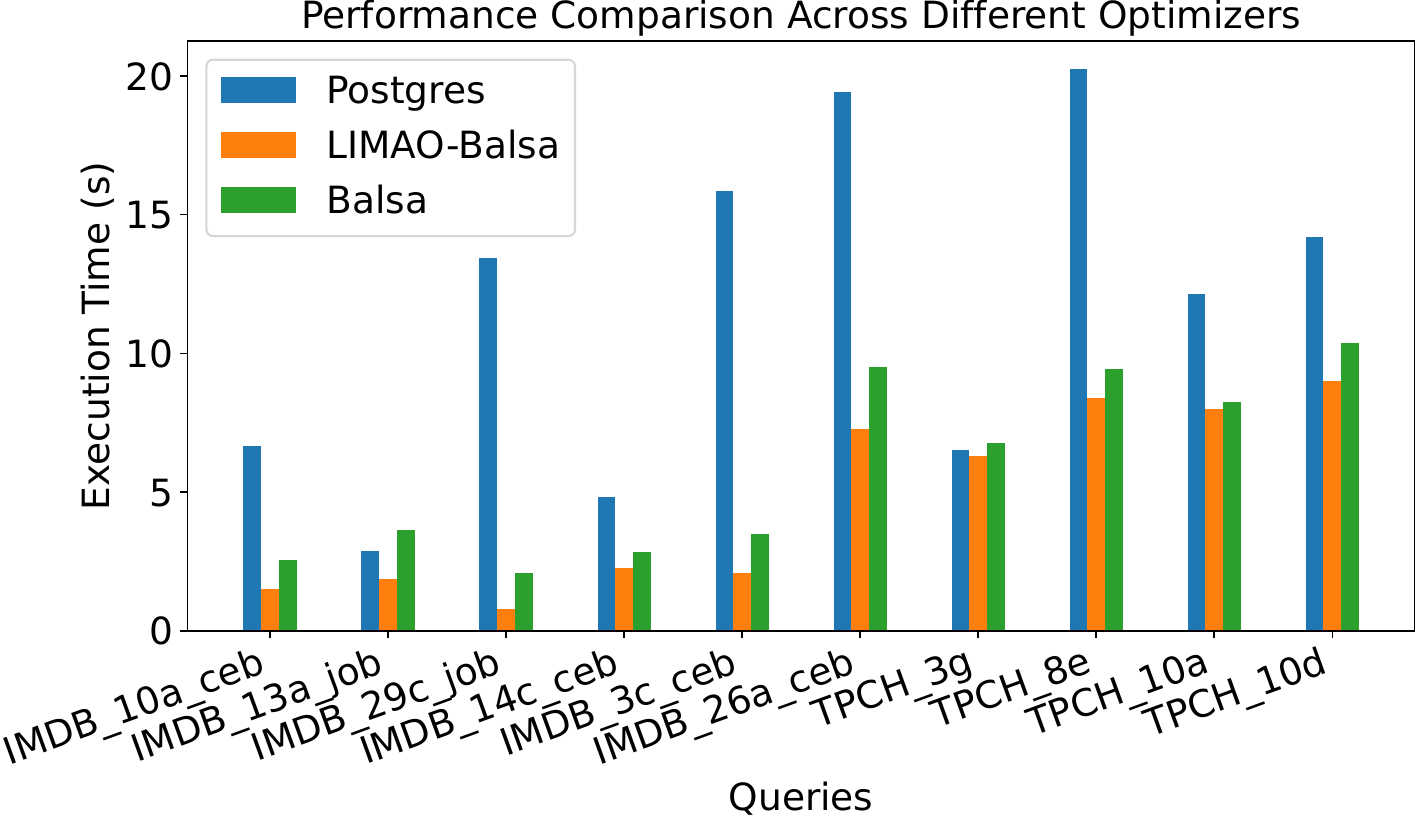}
  \end{center}
  \caption{Sample execution time in simple \textit{Workload Switch}. }
  \label{fig:typical_queries}
  \vspace{-5pt}
\end{figure}

In Figure~\ref{fig:typical_queries}, we provide further insights by analyzing several slow queries: those with execution times exceeding a few seconds, drawn from different workloads to highlight where performance advantages emerge. Consistent with the results in Table~\ref{tab:best_performance}, \balsaimpl generally outperforms Balsa on IMDB queries. TPC-H templates 3, 8, and 10  do not involve complex aggregation functions or nested structures and contain complex join orders. In such cases, a build-from-scratch LQO like Balsa performs well. However, for TPC-H queries with relatively simple join structures, the benefits of using Balsa or \balsaimpl over Postgres are minimal.

\begin{table}[h!]
    \centering
    \small
    \caption{Execution speedup ratio of \balsaimpl vs Balsa in simple \textit{Workload Switch}.}
    \begin{tabular}{cccc}
        \toprule
         Workload & {\balsaimpl} & Balsa & Postgres \\
        \midrule
         \imdbone      & \textbf{2.44} & 1.96 & 1.00 \\
         \imdbtwo      & 2.66          & \textbf{2.68} & 1.00 \\
         \imdbcombined & \textbf{2.59} & 2.40 & 1.00 \\
         \tpchone      & \textbf{1.31} & 1.30 & 1.00 \\
         \tpchtwo      & 0.76          & 0.75 & \textbf{1.00} \\
         \tpchcombined & \textbf{1.11} & 1.10 & 1.00 \\
        \bottomrule
    \end{tabular}
    \label{tab:best_performance}
    \vspace{-5pt}
\end{table}

\vspace{-8pt}

\subsubsection{Challenging Periodic Workload Switch}
\label{subsubsection: challenging situation}

Here, we move to a more complex scenario, where we alternate between {\imdbone}, {\imdbtwo}, and \imdbthree\ every 5 iterations, over a total of 120 iterations, and evaluate {\balsaimpl} against Balsa.
Table~\ref{tab:threeswitch} shows the overall speedup and the variance of plan execution times across iterations. \balsaimpl consistently outperforms Balsa in both metrics, indicating that it produces faster and more stable query plans. Specifically, {\balsaimpl} requires as little as 30\% of Postgres’s execution time and 60\% of Balsa’s, while achieving up to 8.8$\times$ greater stability compared to Balsa. As the environment becomes increasingly dynamic, the stability of \balsaimpl becomes more prominent, thanks to its episode-based training mechanism, which promptly trains the model to prevent generations of bad plans.

\begin{table}[h!]
    \centering
    \small
    \caption{{Speedup ratio and variance of execution time of \balsaimpl} vs Balsa in challenging \textit{Workload Switch}.}
    \begin{tabular}{cccc}
        \toprule
         Metric & {\balsaimpl} & Balsa & Postgres \\
        \midrule
         \imdbone\ Speedup & \textbf{2.49} & 1.53 & 1.00 \\
         \imdbtwo\ Speedup & \textbf{4.08} & 2.92 & 1.00 \\
         \imdbthree\ Speedup & \textbf{1.53} & 1.31 & 1.00 \\
         Combined Speedup & \textbf{2.90} & 2.07 & 1.00\\ 
         \imdbone\ Variance & \textbf{$ \boldsymbol{4.73 \times 10^5}$} &$4.20 \times 10^6$& NA \\
         \imdbtwo\ Variance &  $\boldsymbol{1.18 \times 10^6}$&$1.05 \times 10^7$& NA \\
         \imdbthree\ Variance & $\boldsymbol{5.21 \times 10^5}$   &$2.29 \times 10^6$& NA \\
        \bottomrule
    \end{tabular}
    
    \label{tab:threeswitch}
    \vspace{-5pt}
\end{table}

\begin{figure}
  \begin{center}
\includegraphics[width=0.45\textwidth]{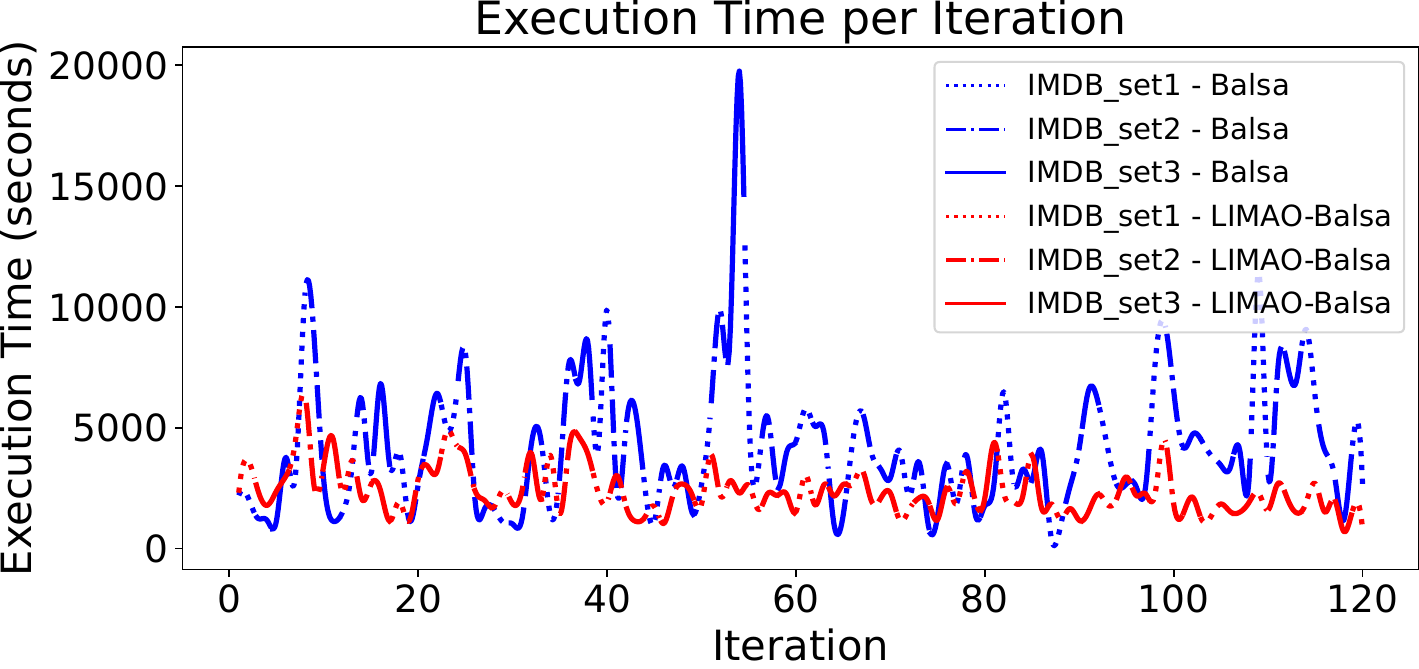}
  \end{center}
  \caption{Execution time throughout 120 iterations in the challenging \textit{Workload Switch}. }
  \label{fig:three_workloads}
  \vspace{-8pt}
\end{figure}

Figure~\ref{fig:three_workloads} visually depicts the execution time of Balsa and {\balsaimpl}, summed across all training and testing queries within each iteration, with different line segments indicating the active workload at each point. {\balsaimpl} exhibits significantly greater stability and shows steady performance improvements over time. In contrast, Balsa’s performance curve is highly erratic, with frequent spikes that correspond to instances of catastrophic forgetting, where the model generates poor plans upon switching back to a previously seen workload. This figure highlights {\balsaimpl}’s ability to effectively mitigate catastrophic forgetting, demonstrating its robustness and practical viability for real-world dynamic environments.

\subsubsection{Challenging Non-periodic Workload Switch.} \label{subsubsection:non_periodic}
\begin{figure}[h]
 \hspace*{-1.2cm}
\includegraphics[width=0.4\textwidth]  {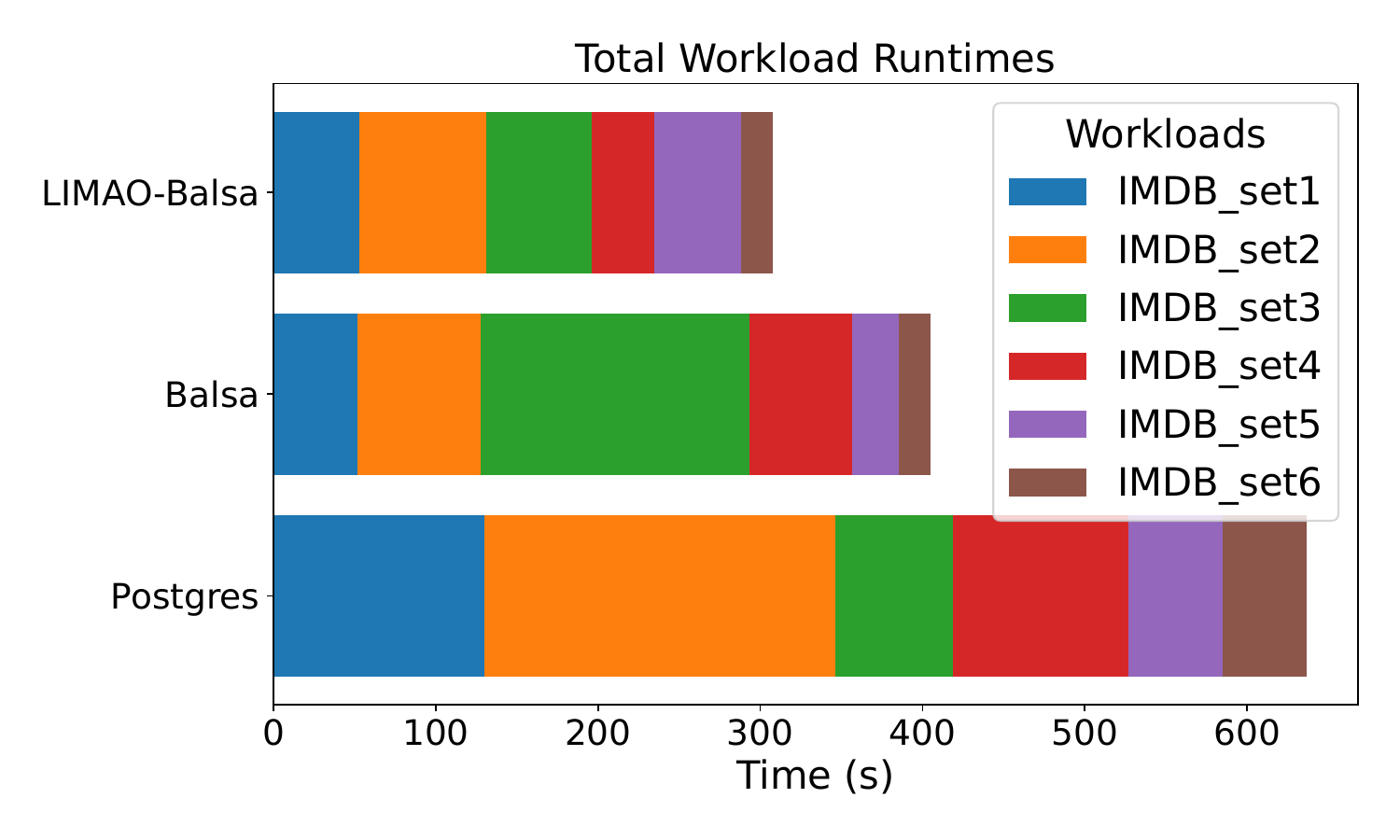}
  \caption{Total optimal execution time of different workloads in non-periodic cross-schema \textit{Workload Switch}. }
  \label{fig:stacked_bar}
  \vspace{-5pt}
\end{figure}

In this experiment, we use all six IMDB workloads, with workloads switching randomly: each executed for 2 to 8 consecutive iterations. To ensure reproducibility, we use fixed random seeds. The total number of iterations is set to 100. As shown in Figure~\ref{fig:stacked_bar}, Balsa requires 405 seconds to complete all workloads optimally, while \balsaimpl achieves the same in just 307 seconds. Balsa encounters 464 timeouts (defined as queries exceeding 60 seconds in execution time), compared to only 219 timeouts with {\balsaimpl}. These results demonstrate that \oursystem significantly improves Balsa’s performance in highly unpredictable, non-periodic environments.

\begin{table}[h!]
    \centering
    \small
    \caption{Speedup ratio of \baoimpl vs Bao in non-periodic \textit{Workload Switch} for \baoimpl experiment.}
    \begin{tabular}{cccc}
        \toprule
         Workload & {\baoimpl} & Bao & Postgres \\
        \midrule
         IMDB\_set1       & \textbf{1.17} & \textbf{1.17} & 1.00 \\
         IMDB\_set2       & \textbf{1.48}          & 1.25 & 1.00 \\
         IMDB\_Combined   & \textbf{1.34}          & 1.22 & 1.00 \\
         TPC-H\_set1      & 1.98 & \textbf{2.04} & 1.00 \\
         TPC-H\_set2      & \textbf{3.20} & \textbf{3.20} & 1.00 \\
         TPC-H\_set3      & \textbf{1.97} & 1.58 & 1.00 \\
         TPC-H\_Combined  & \textbf{2.15} & 1.91 & 1.00 \\
All 5 Workloads  & \textbf{1.85} & 1.66 & 1.00 \\
        \bottomrule
    \end{tabular}
    \label{tab:best_performance_bao}
    \vspace{-5pt}
\end{table}

\vspace{-8pt}

\subsubsection{Non-periodic \textit{Workload Switch} with Schema Changes.}
\label{subsubsection:non-periodic-bao}
In this section, we evaluate \baoimpl under a non-periodic \textit{Workload Switch} scenario involving schema changes. We randomly switch between five workloads: \imdbone, \imdbtwo, \tpchone, \tpchtwo, and \tpchthree, where each workload is executed consecutively for 2 to 10 iterations. Table~\ref{tab:best_performance_bao} reports the speedup ratios of Bao and \baoimpl under this dynamic setting. Overall, \baoimpl shows either promising improvements or performance comparable to Bao across most workloads. \baoimpl achieves an average speedup of 1.85× over Postgres, compared to 1.66× achieved by Bao. Both Bao and \baoimpl perform particularly well on the TPC-H benchmarks, which can be attributed to Bao’s hint-based design that is especially effective for handling simpler, join-naive queries.

\subsubsection{Performance of \balsaimpl and Balsa in All Situations}\label{subsubsection:performance comparison over iterations on four situations}

\begin{figure*}

  \begin{subfigure}[b]{0.2\linewidth}
    \begin{center}
    \includegraphics[width=1.7in]{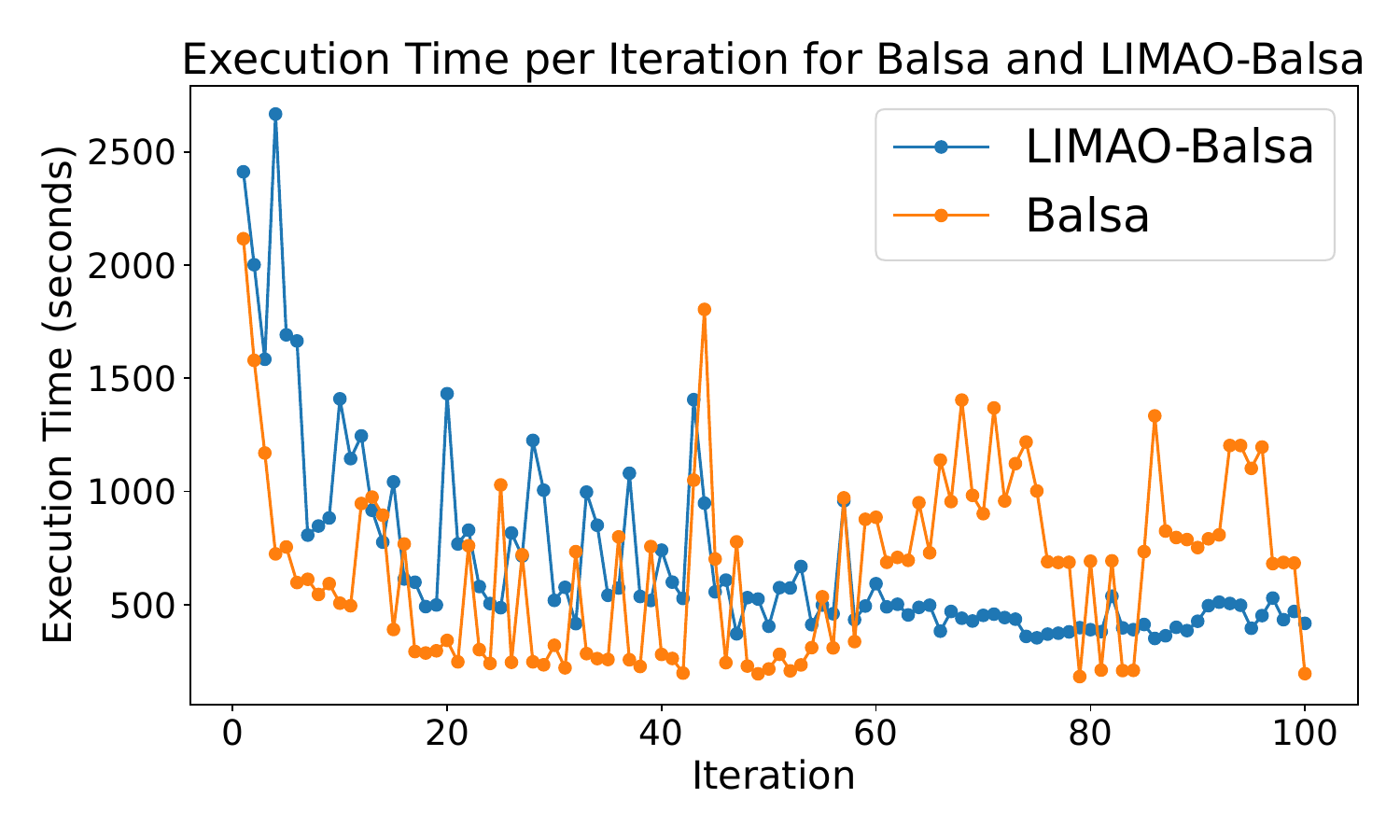}
    \end{center}
    \subcaption{IMDB \textit{Static Workload}.}
    \label{fig:performance_line_IMDB_static}
  \end{subfigure}
  \hspace{0.5cm}
  \begin{subfigure}[b]{0.2\linewidth}
    \begin{center}
    \includegraphics[width=1.7in]{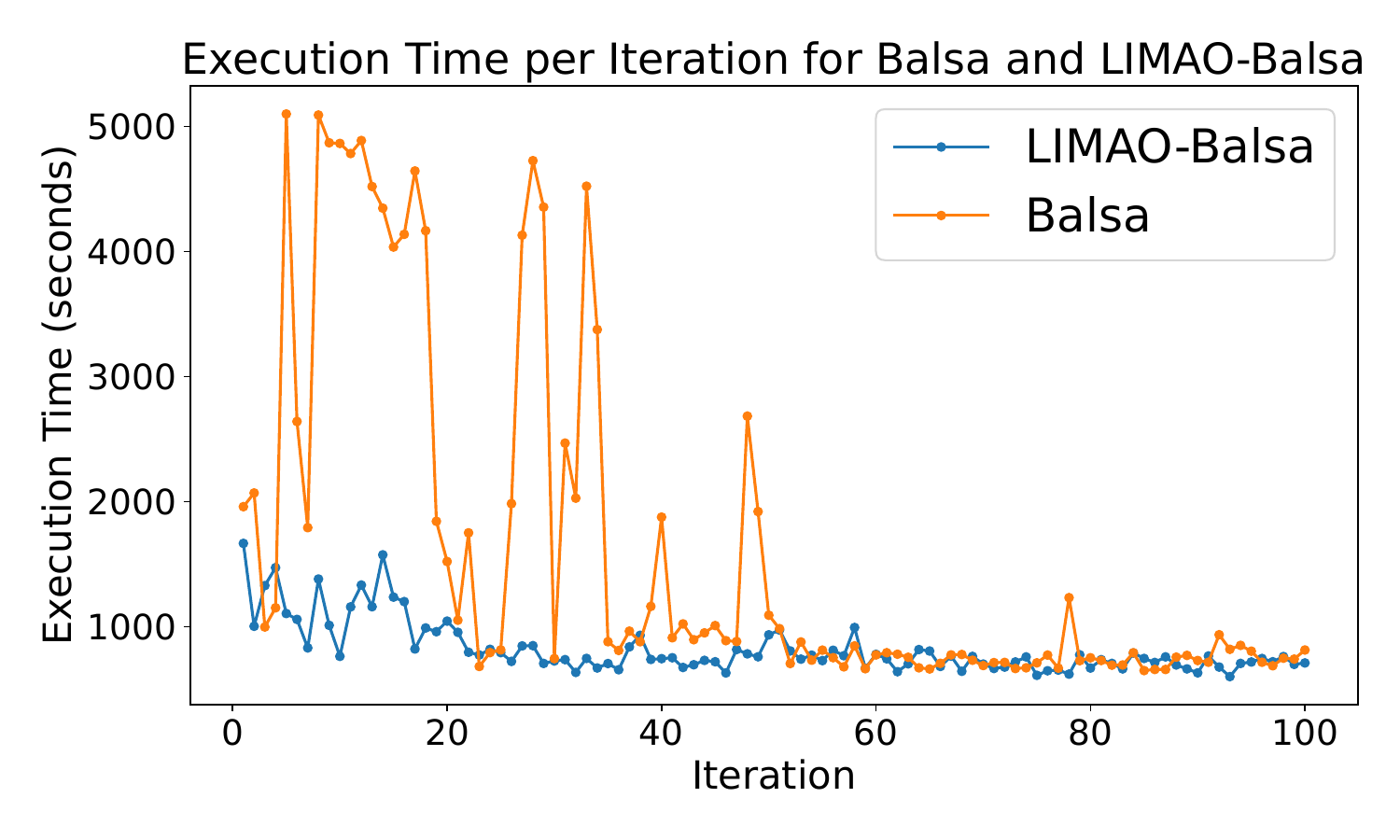}
    \end{center}
    \subcaption{TPC-H \textit{Static Workload}.}
    \label{fig:performance_line_TPCH_static}
  \end{subfigure}
  \hspace{0.5cm}
  \begin{subfigure}[b]{0.2\linewidth}
    \begin{center}
    \includegraphics[width=1.7in]{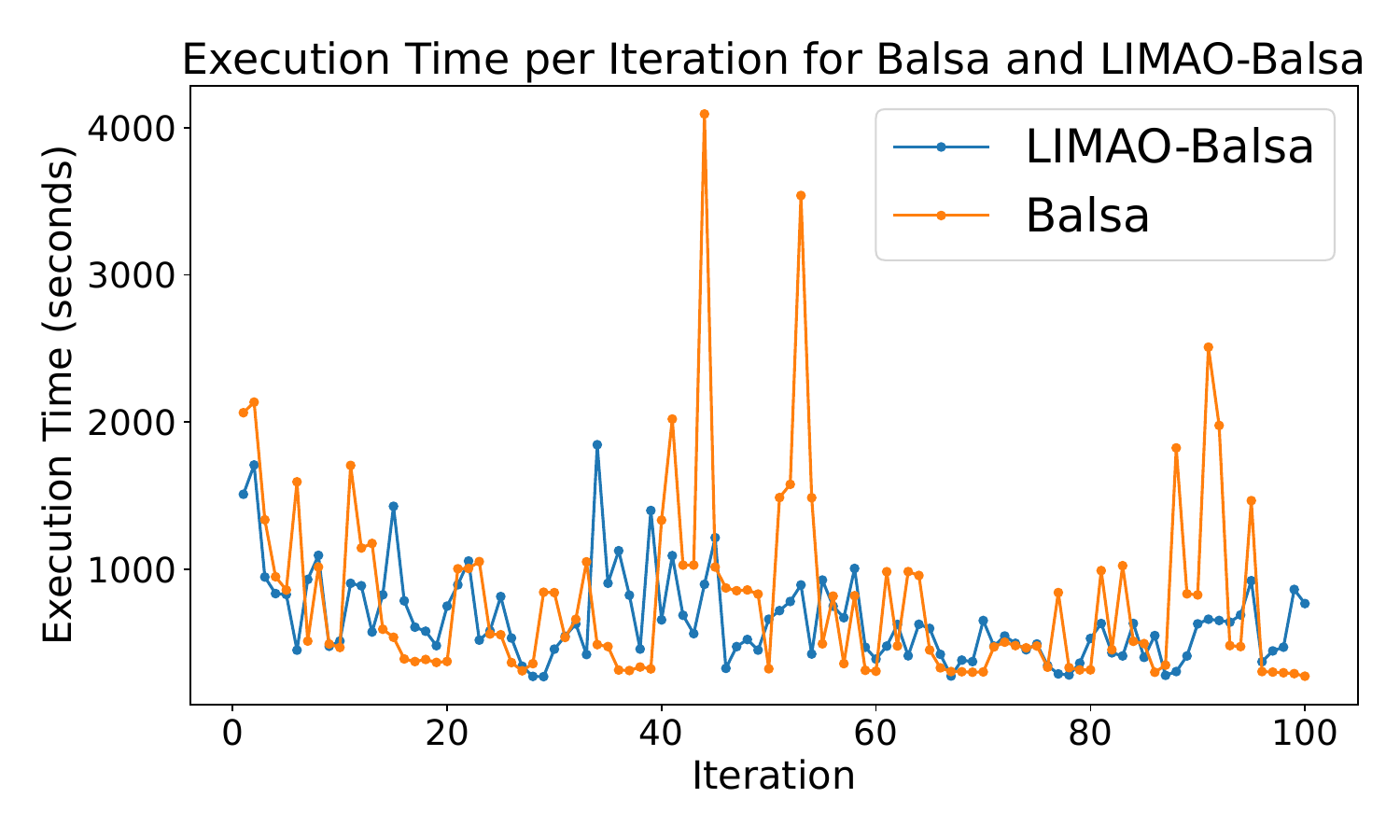}
    \end{center}
    \subcaption{IMDB \textit{Volume Switch}.}
    \label{fig:performance_line_IMDB_switch_db}
  \end{subfigure}
  \hspace{0.5cm}
  \begin{subfigure}[b]{0.2\linewidth}
    \begin{center}
    \includegraphics[width=1.7in]{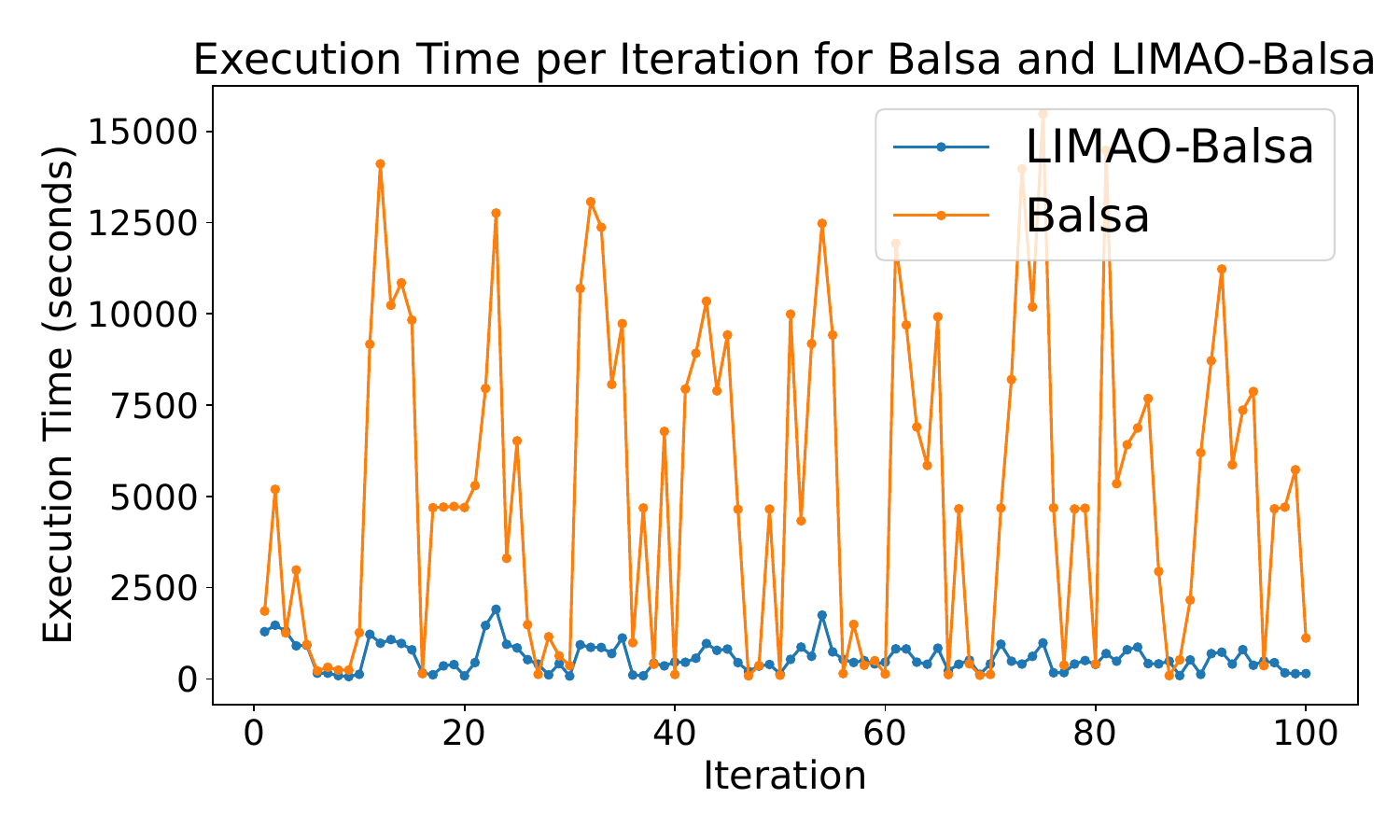}
    \end{center}
    \subcaption{TPC-H \textit{Volume Switch}.}
    \label{fig:performance_line_TPCH_switch_db}

  \end{subfigure}

  \begin{subfigure}[b]{0.2\linewidth}
    \begin{center}
    \includegraphics[width=1.7in]{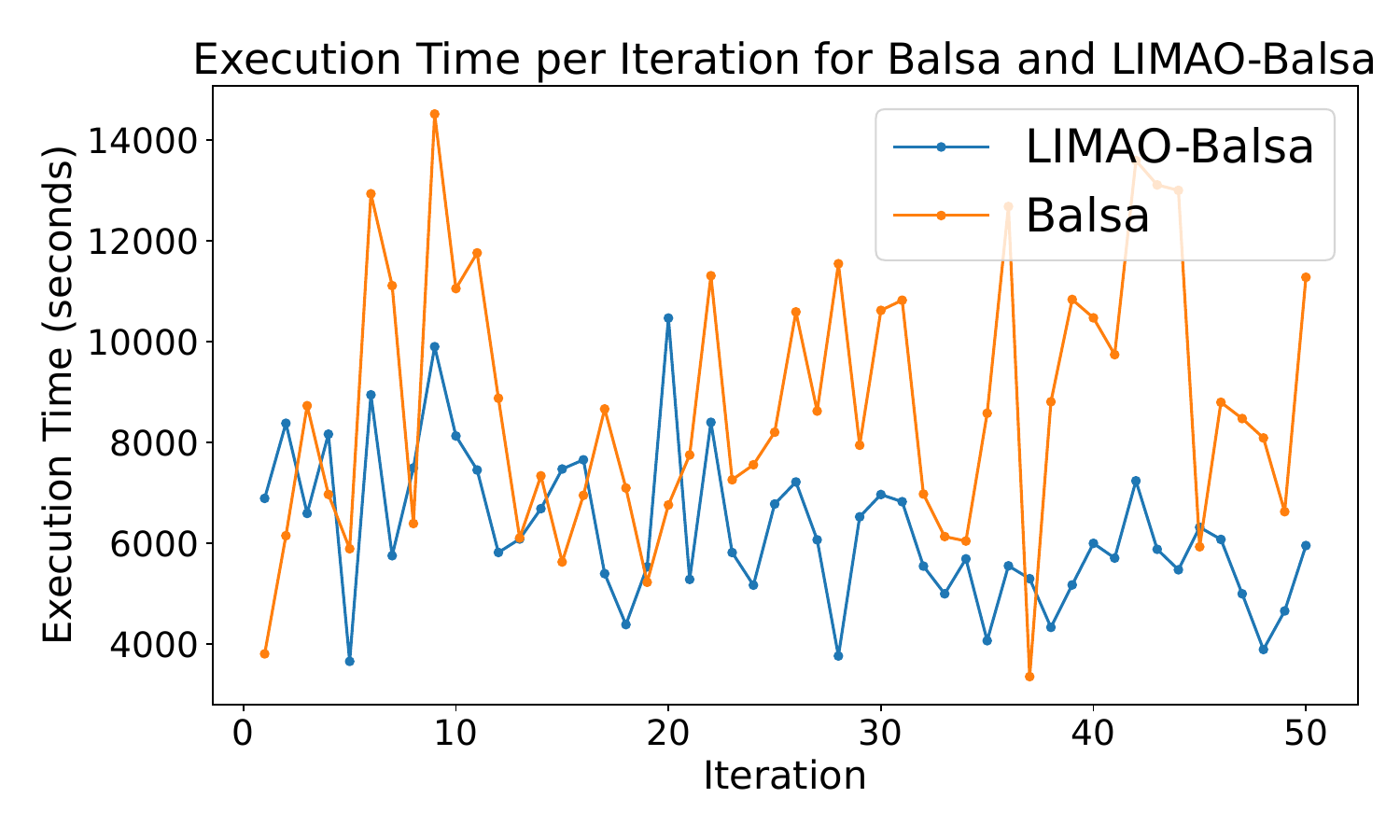}
    \end{center}
    \subcaption{IMDB \textit{Workload Switch}.}
    \label{fig:performance_line_IMDB_switch_workload}

  \end{subfigure}
  \hspace{0.5cm}
  \begin{subfigure}[b]{0.2\linewidth}
    \begin{center}
    \includegraphics[width=1.7in]{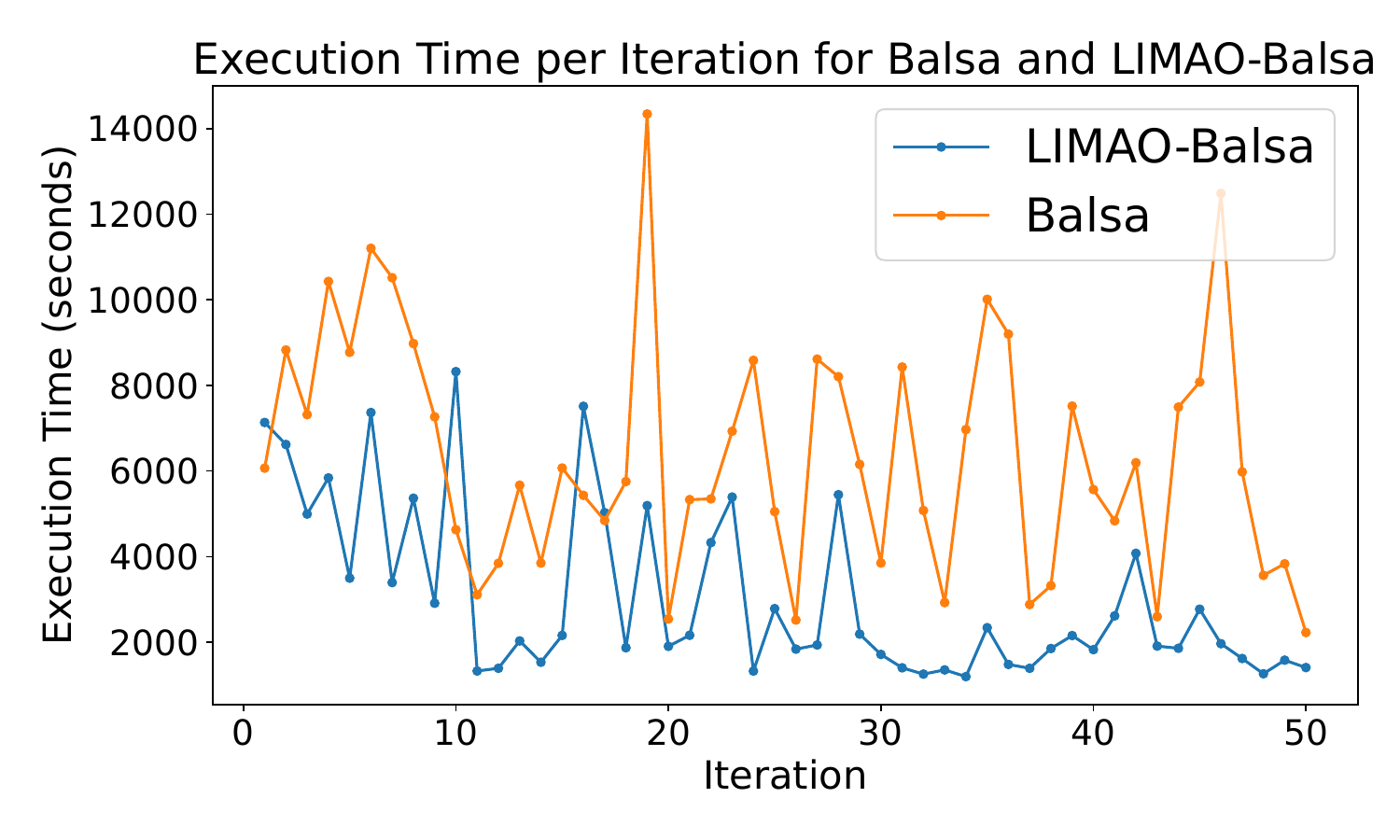}
    \end{center}
    \subcaption{TPC-H \textit{Workload Switch}.}
    \label{fig:performance_line_TPCH_switch_workload}
  \end{subfigure}
  \hspace{0.5cm}
  \begin{subfigure}[b]{0.2\linewidth}
    \begin{center}
    \includegraphics[width=1.7in]{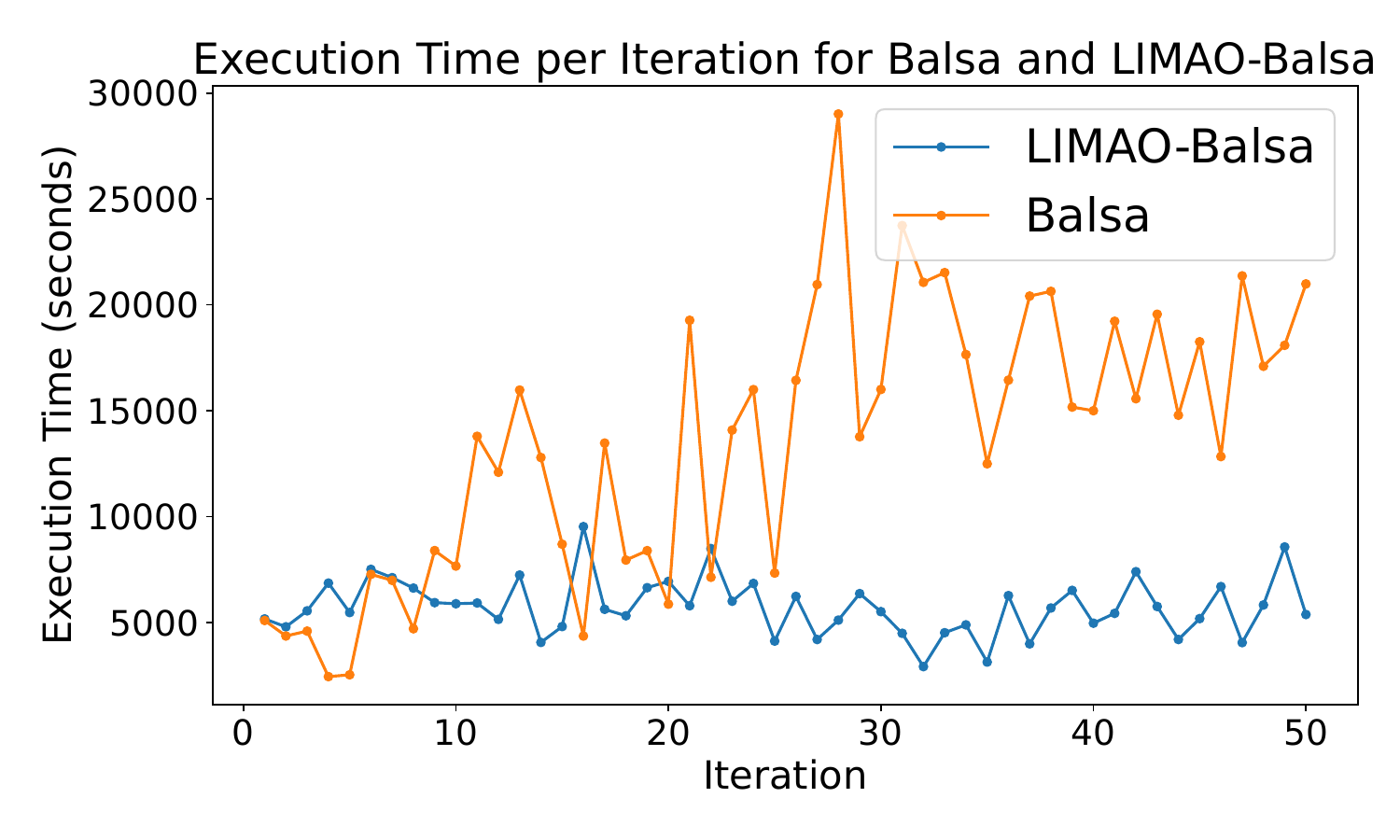}
    \end{center}
    \subcaption{IMDB  \textit{Both Switch}.}
    \label{fig:performance_line_IMDB_switch_both}
  \end{subfigure}
  \hspace{0.5cm}
  \begin{subfigure}[b]{0.2\linewidth}
     \begin{center}
    \includegraphics[width=1.7in]{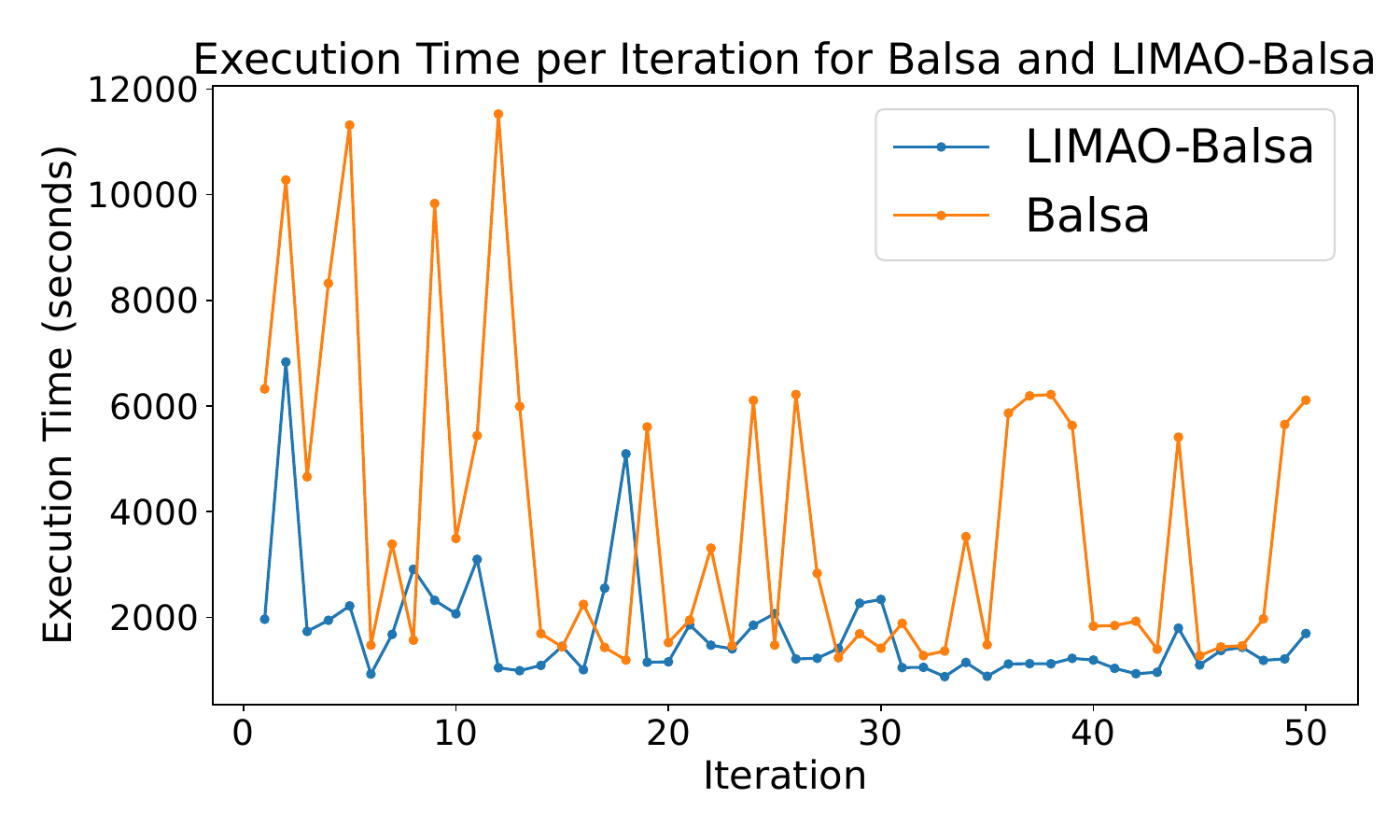}
    \end{center}
    \subcaption{TPC-H \textit{Both Switch}.}
    \label{fig:performance_line_TPCH_switch_both}
  \end{subfigure}
  
  \caption{Performance of Balsa and \balsaimpl throughout iterations in all 8 types of dynamic environments.}
  \label{fig:performance_main_compare}
  \vspace{-5pt}
\end{figure*}

Figure~\ref{fig:performance_main_compare} presents the execution time of static workloads and three types of dynamic switching scenarios across iterations, highlighting the overall performance stability of \balsaimpl and Balsa. While both achieve comparable overall performance, {\balsaimpl} demonstrates a significantly more stable execution pattern, whereas Balsa suffers from noticeable fluctuations.

In the \textit{Static} setting (Figures~\ref{fig:performance_line_IMDB_static} and~\ref{fig:performance_line_TPCH_static}), \balsaimpl performs comparably to Balsa, but with consistently lower execution time in most iterations. Notably, in the TPC-H \textit{Static} scenario, \balsaimpl exhibits much lower execution time at the beginning, whereas Balsa produces a continuous segment of poorly performing plans. This trend indicates that \balsaimpl's episodic training mechanism helps prevent the repeated generation of suboptimal plans, even in static environments. In the \textit{Both Switch} scenarios (Figures~\ref{fig:performance_line_IMDB_switch_both} and~\ref{fig:performance_line_TPCH_switch_both}), Balsa's performance is highly unstable, especially in the IMDB environment, where performance even deteriorates over time. In contrast, \balsaimpl maintains low execution latency and stable performance across all iterations. These results demonstrate that {\balsaimpl} is more robust than Balsa across both static and dynamic scenarios. This improvement can be attributed to the \textit{lifelong} {\lcp}, which stabilizes the training process and reduces the occurrence of performance spikes—an issue commonly observed in conventional reinforcement learning approaches.

\subsection{Micro Benchmarks}
\label{sec:micro_bench}
\begin{figure}[t]
  \centering
  \begin{subfigure}[b]{0.45\linewidth}
    \centering
    \includegraphics[width=1.6in]{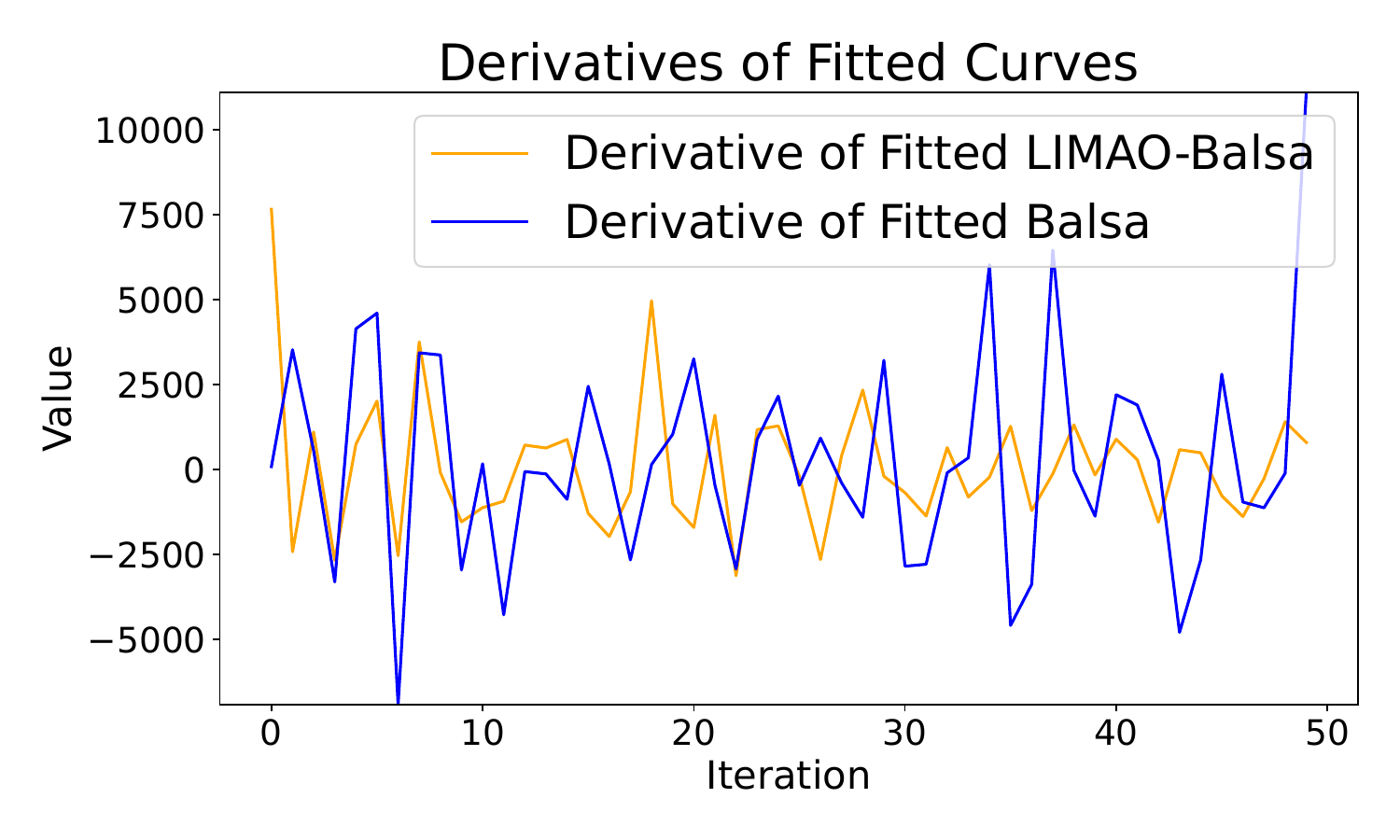}
    \subcaption{IMDB \textit{Workload Switch}.}
  \end{subfigure}
  \hspace{0.5cm}
    \begin{subfigure}[b]{0.45\linewidth}
    \centering
    \includegraphics[width=1.6in]{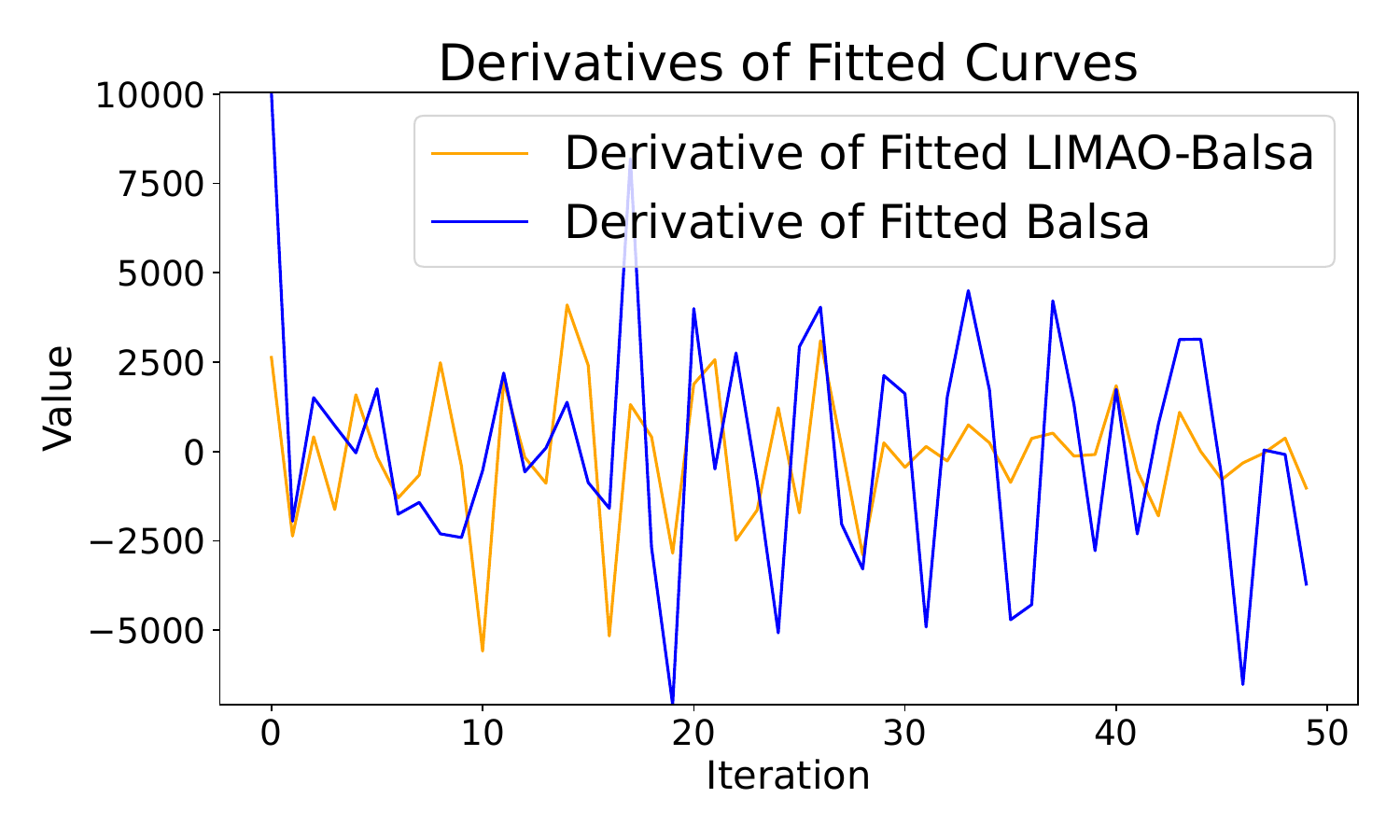}
    \subcaption{TPC-H \textit{Workload Switch}.}
  \end{subfigure}
  \vspace{0.2cm}
    \vspace{0.2cm}
    \begin{subfigure}[b]{0.45\linewidth}
    \centering
    \includegraphics[width=1.6in]{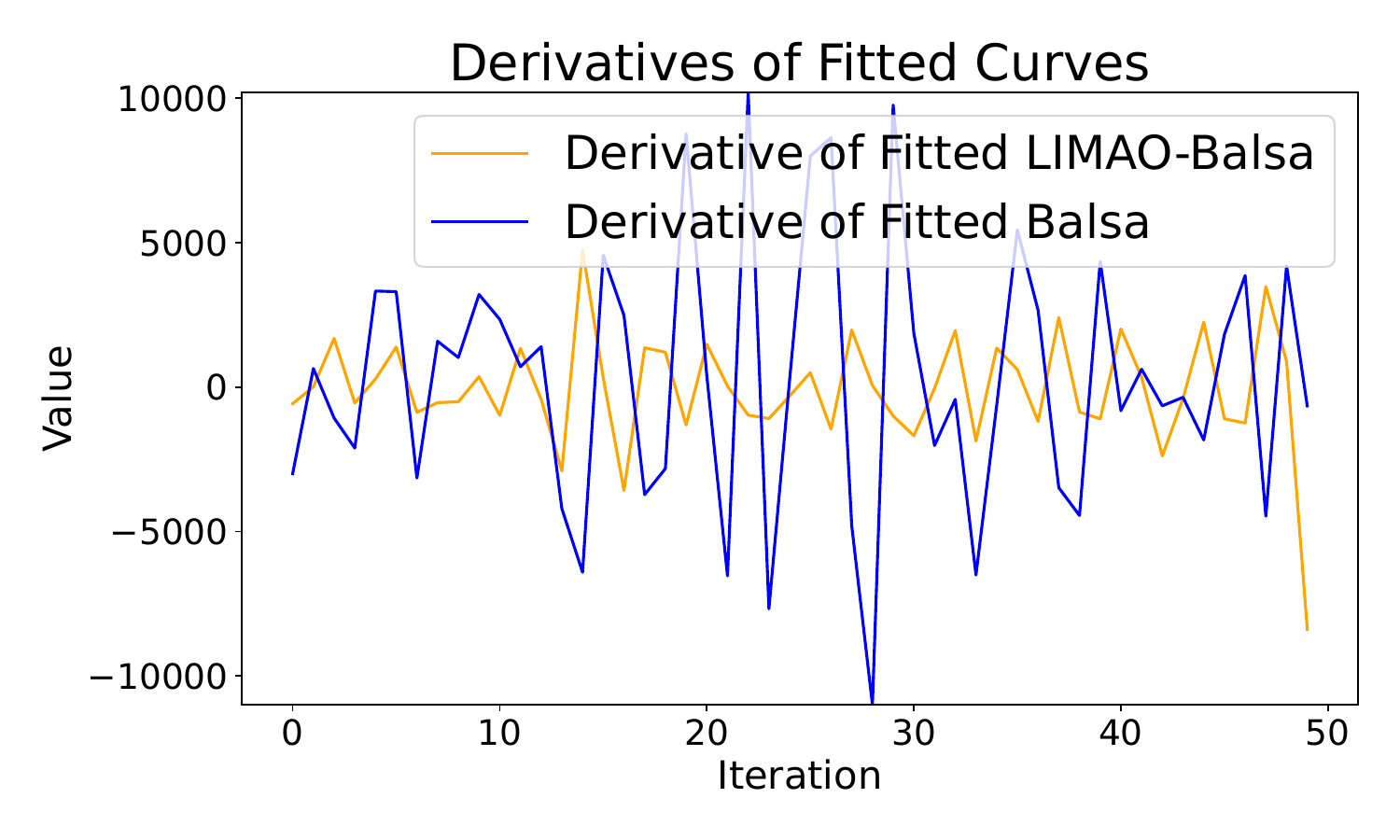}
    \subcaption{IMDB \textit{Both Switch}.}
  \end{subfigure}
  \hspace{0.5cm}
  \begin{subfigure}[b]{0.45\linewidth}
    \centering
    \includegraphics[width=1.6in]{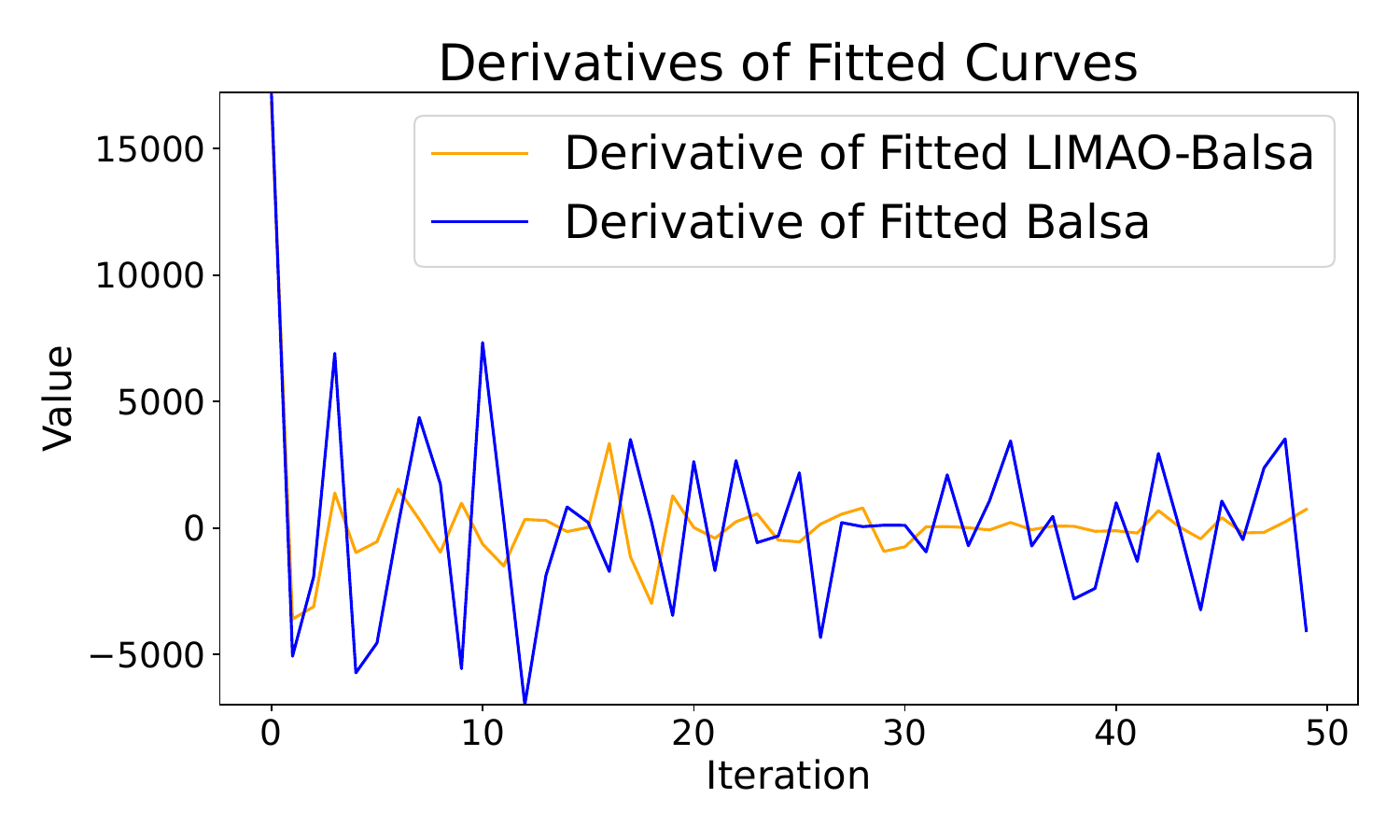}
    \subcaption{TPC-H \textit{Both Switch}.}
  \end{subfigure}
  \caption{Derivatives of execution time from each iteration.}
  \label{fig:derivative_graph}
  \vspace{-10pt}
\end{figure}

\vspace{-2pt}
\subsubsection{Performance Stability}
\label{sec:micro_benchmark_performance_stability}
 During training iterations, {\balsaimpl} demonstrates notable performance stability. In this experiment, we collect the total execution time at each iteration, smooth the resulting time series into a continuous curve, and compute its derivative to assess stability. Figure~\ref{fig:derivative_graph} shows the derivatives of the fitted curves for both Balsa and {\balsaimpl}. In dynamic environments, Balsa exhibits substantial instability, as reflected by its large and highly variable derivative values. Such instability often results in severe performance degradation, as discussed in detail in Section~\ref{section:Disastrous_plans}. Moreover, consistently lower derivative values suggest that a model is converging, as the differences in execution time between consecutive iterations become minimal. For instance, in the IMDB \textit{Workload Switch} scenario, {\balsaimpl}'s derivative values remain below 1500 after 30 iterations, compared to 11,000 for Balsa. Similarly, in the TPC-H \textit{Both Switch} scenario, {\balsaimpl}'s values stay below 1000 after 20 iterations, whereas Balsa's rise to 4400. These observations indicate that {\balsaimpl} likely converges significantly faster, potentially requiring only slightly more than half the number of iterations compared to Balsa. Table~\ref{tab:variances} further supports this observation by presenting the variance in execution time across different scenarios. In every case, {\balsaimpl} exhibits lower variance than Balsa, often by orders of magnitude. Notably, in the IMDB \textit{Both Switch} scenario, {\balsaimpl} achieves 22$\times$ greater stability than Balsa. These results highlight that \oursystem substantially improves Balsa’s robustness particularly when both query workloads and data distributions shift.

\begin{table}[h!]
    \centering
    \small
    \caption{Variances for Balsa and {\balsaimpl}. The results are based on each iteration's combined execution time.}
    \begin{tabular}{cccc}
        \toprule
         Workload & pattern & {\balsaimpl} & Balsa  \\
        \midrule
         IMDB & \textit{Both Switch} &  $\boldsymbol{1.77 \times 10^6}$ & $4.00 \times 10^7$ \\
         IMDB &  \textit{Workload Switch} &  $\boldsymbol{2.23 \times 10^6}$ & $6.96 \times 10^6$ \\
         IMDB &  \textit{Volume Switch} &  $\boldsymbol{9.43 \times 10^4}$ & $4.32 \times 10^5$ \\
         TPC-H &  \textit{Both Switch} &  $\boldsymbol{1.10 \times 10^6}$ & $8.30 \times 10^6$ \\
         TPC-H & \textit{Workload Switch} &  $\boldsymbol{3.93 \times 10^6}$ & $7.70 \times 10^6$ \\
         TPC-H &  \textit{Volume Switch} &  $\boldsymbol{1.44 \times 10^5}$ & $1.88 \times 10^7$ \\
        \bottomrule
    \end{tabular}
    
    \label{tab:variances}
    \vspace{-5pt}
\end{table}

\subsubsection{Choice of Module Hub's Length} \label{subsubsection:hub_length}To evaluate the impact of different Module Hub sizes, we test three settings on the IMDB workload by varying the number of modules assigned to the HJ, NL, and OTH hubs. In setting S1, we assign 1 module each to HJ, NL, and OTH. In setting S2, we use 2 modules for HJ, 3 for NL, and 1 for OTH. In setting S3, we assign 3 modules to each of HJ, NL, and OTH. These three settings are evaluated over 20 iterations in the IMDB  \textit{Workload Switch}  scenario to compare their minimum execution time per iteration. The results for S1, S2, and S3 are $4067s$, $3652s$, and $4826s$, respectively (S2 has the best performance). This result highlights how the choice of Module Hub size can significantly influence the performance of \oursystem. In the IMDB workload, NL joins tend to exhibit more complex behavior than HJ and OTH operators, so allocating more modules to NL tasks improves model performance. However, assigning too many modules can lead to underfitting, as the number of training tasks per module becomes insufficient, ultimately degrading overall performance.

\sloppy\subsubsection{Ablation Study.}\label{subsubsection:ablation_study} 
\begin{figure}[t]
  \setlength{\abovecaptionskip}{2pt} 
  \setlength{\belowcaptionskip}{-2pt}
  \centering
  \begin{subfigure}[b]{0.45\linewidth}
    
    \centering
    \includegraphics[width=1.7in]{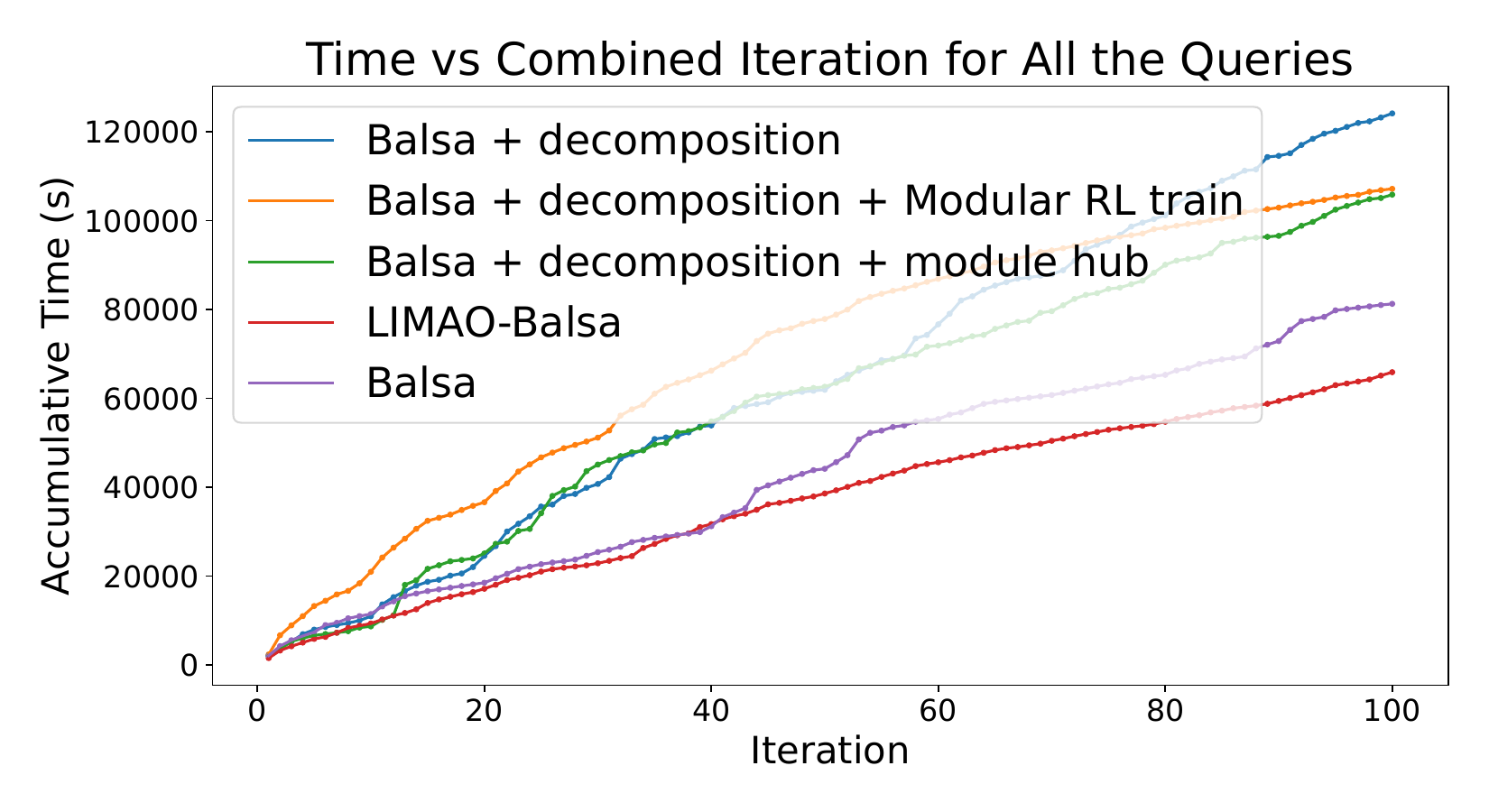}
    \subcaption{IMDB \textit{Volume Switch}.}
    \label{fig:ablation_study_a}
  \end{subfigure}
  \hspace{0.5cm}
    \begin{subfigure}[b]{0.45\linewidth}
    \centering
    \includegraphics[width=1.7in]{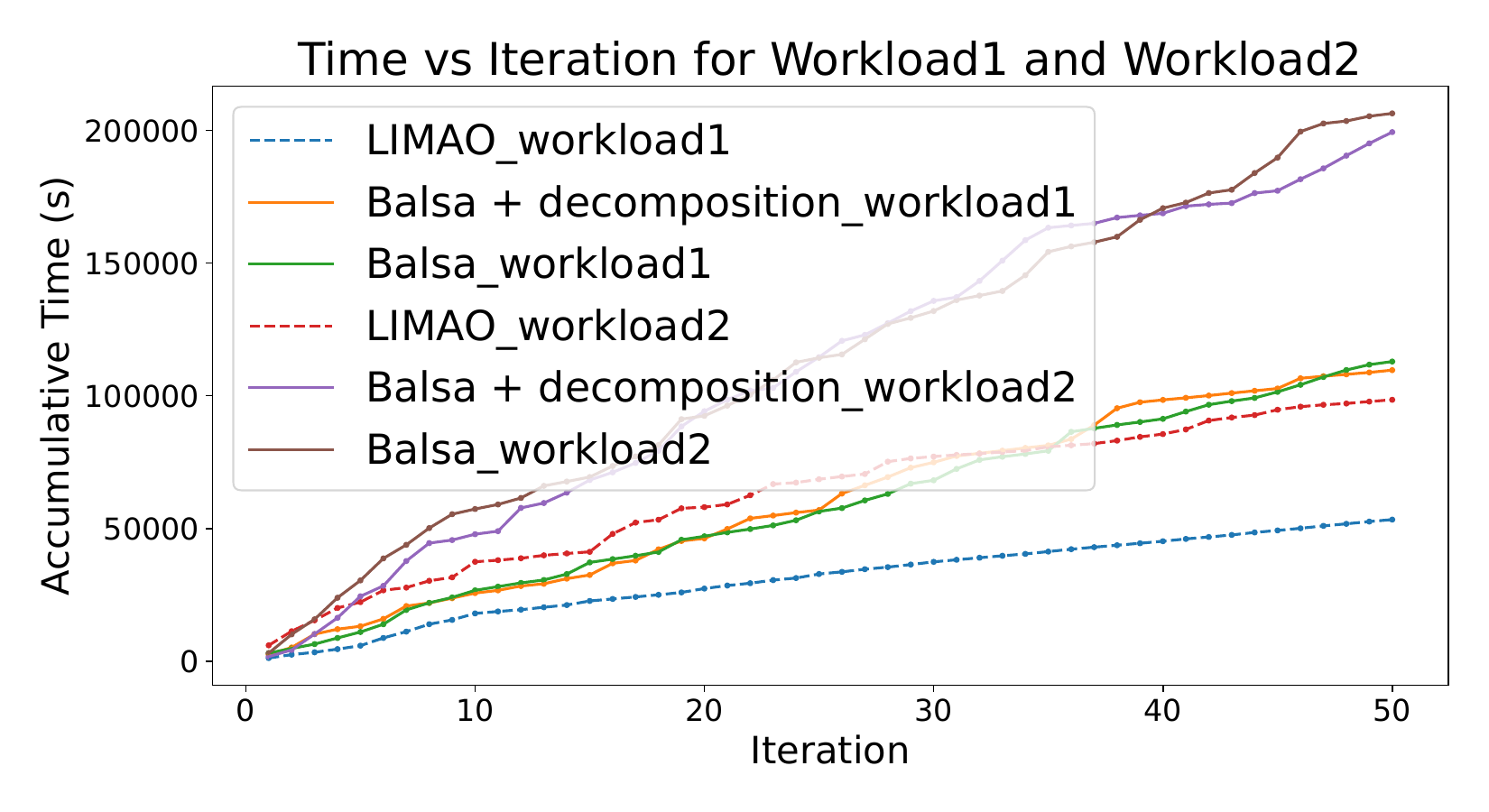}
    \subcaption{TPC-H \textit{Workload Switch}.}
        \label{fig:ablation_study_b}

  \end{subfigure}
  \vspace{0.2cm}
    \begin{subfigure}[b]{0.45\linewidth}
    \centering
    \includegraphics[width=1.7in]{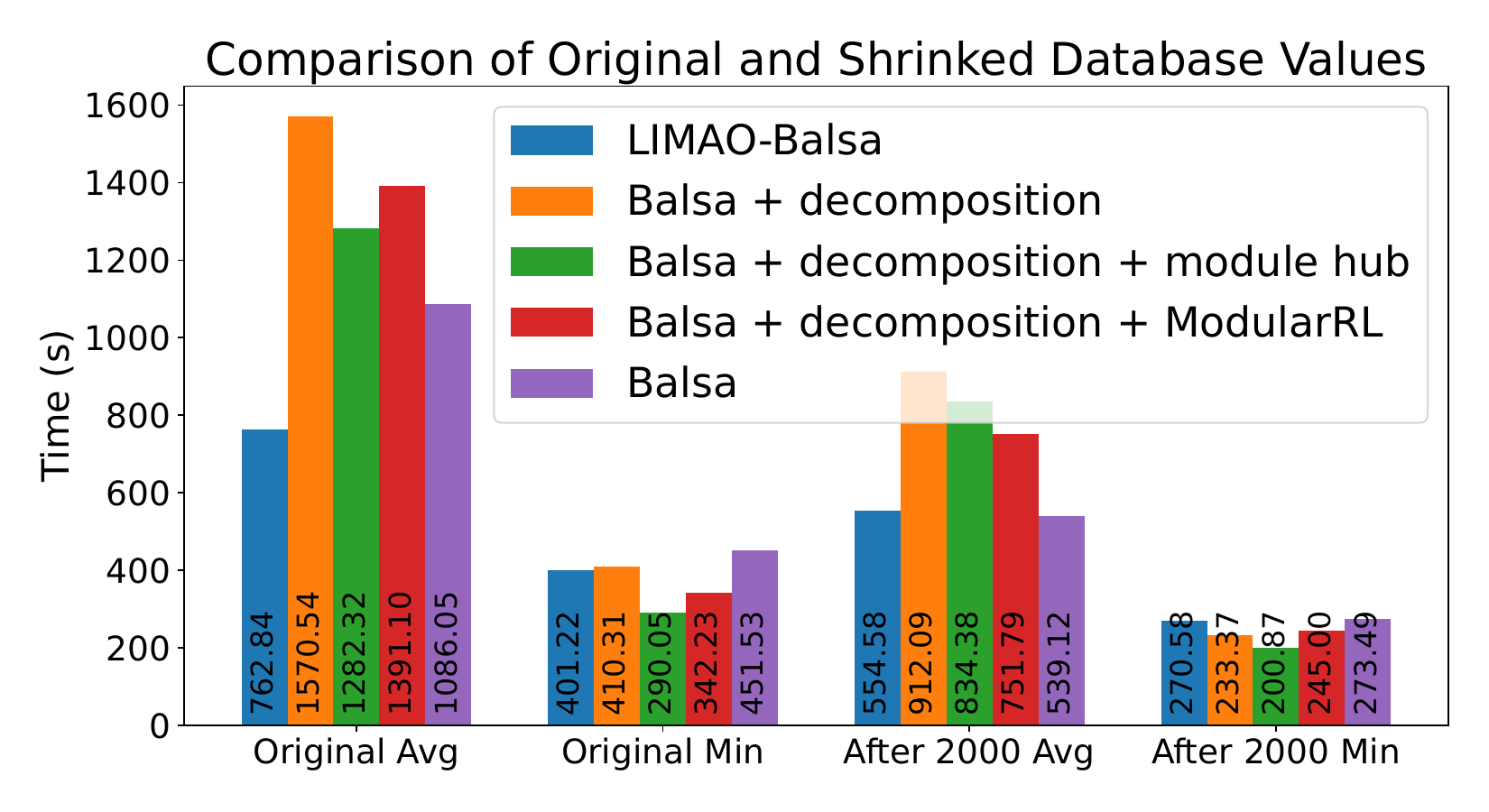}
    \subcaption{IMDB Average and MIN Running Time.}
        \label{fig:ablation_study_c}

  \end{subfigure}
  \hspace{0.5cm}
  \begin{subfigure}[b]{0.45\linewidth}
    \centering
    \includegraphics[width=1.7in]{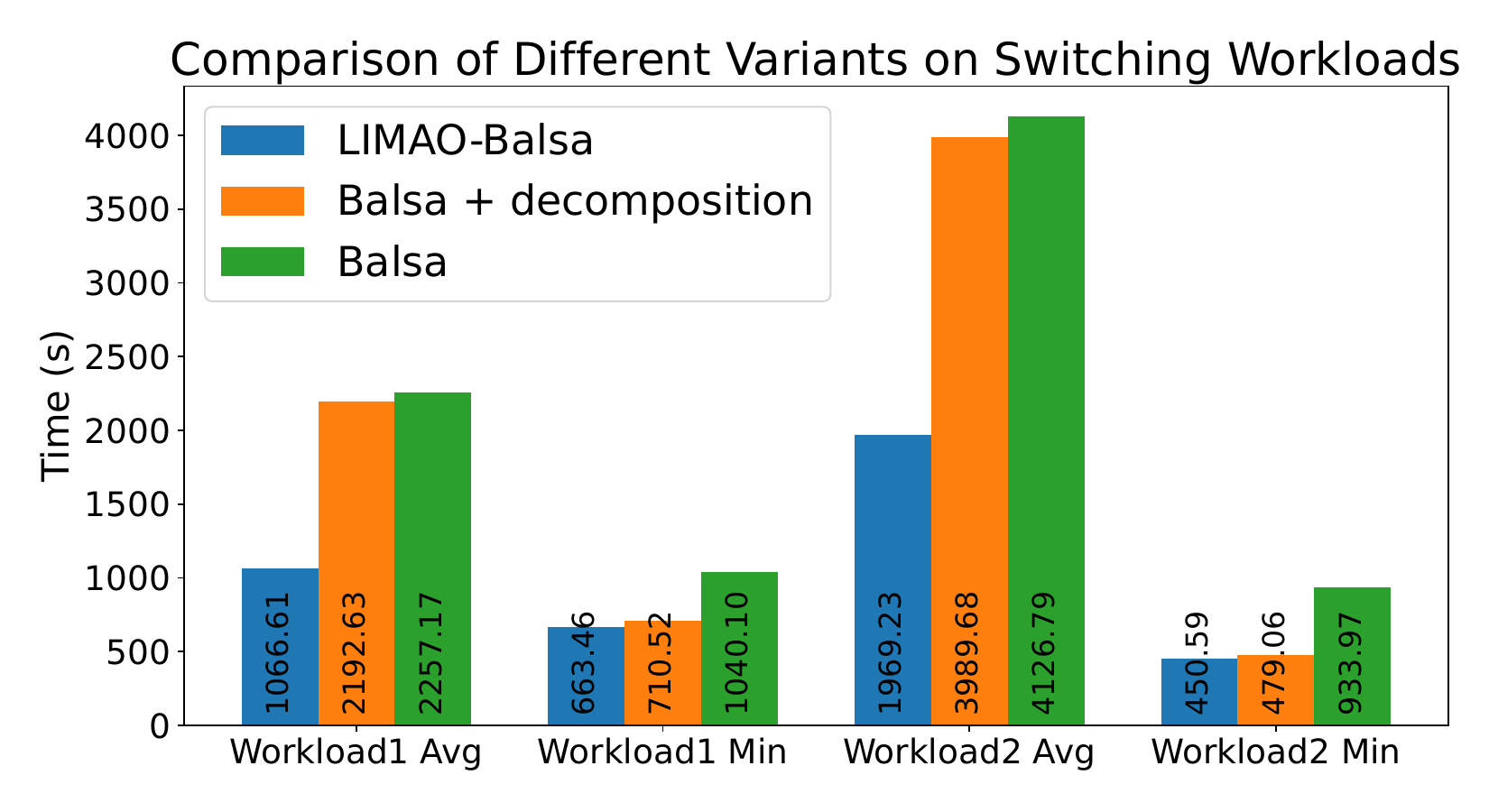}
    \subcaption{TPC-H Average and MIN Running Time.}
        \label{fig:ablation_study_d}

  \end{subfigure}
  
  \caption{Accumulated, average, and minimum execution performance of {\balsaimpl} variations. }
  \label{fig:ablation_study}
  \vspace{-8pt}
\end{figure}

To evaluate the impact of different components in \oursystem, we conduct ablation studies in the IMDB \textit{Volume Switch} and TPC-H \textit{Workload Switch} environments. For the IMDB workload, we compare the following variations: (1) Balsa, (2) \textsl{Balsa + decomposition}, (3) \textit{Balsa + decomposition + Module Hub}, (4) \textit{Balsa + decomposition + Modular RL training}, and (5) the full {\balsaimpl}. As described in Section~\ref{exp_setup}, each Module Hub in the TPC-H workload is limited to one module due to the simpler join patterns characteristic of TPC-H. Accordingly, we evaluate only three configurations for TPC-H: Balsa, \textsl{Balsa + decomposition}, and \textit{Balsa + decomposition + Modular RL training} (i.e., {\balsaimpl} without the Module Hub component).
Figure~\ref{fig:ablation_study} shows the performance of Balsa and the various \balsaimpl\ variants. As shown in Figures~\ref{fig:ablation_study_a} and~\ref{fig:ablation_study_b}, {\balsaimpl} achieves the lowest cumulative execution time in both IMDB and TPC-H workloads, demonstrating its robustness and effectiveness in dynamic environments. In Figures~\ref{fig:ablation_study_c} and~\ref{fig:ablation_study_d}, we observe that while some variants occasionally discover better individual query plans than Balsa and \balsaimpl, they suffer from higher average execution times overall. This is primarily because, without the use of a replay buffer, these variants tend to overfit to a narrow subset of queries. A specific module combination may be trained on only a limited set of queries and thus fail to generalize when presented with new ones, resulting in performance collapse. Similarly, using decomposition alone can lead to overfitting and poor generalization. This issue is particularly evident in the variant combining tree decomposition with Module Hubs. In contrast, {\balsaimpl} achieves an optimal balance between short-term peak performance and long-term stability.

\vspace{-8pt}
\subsubsection{Bad Performance Plans} 
\label{section:Disastrous_plans}
A query plan is considered bad if its execution time exceeds 512 seconds. Table~\ref{tab:timeout_number} reports the number of such timeouts across different scenarios for both Balsa and {\balsaimpl}. Balsa experiences significantly more timeouts during model training, while {\balsaimpl} effectively avoids them. When {\balsaimpl} encounters a bad plan, it is less likely to generate a similar one in subsequent iterations, due to its episodic training mechanism that enables prompt model tuning. This enhanced stability stems from the \textit{lifelong} {\lcp} training paradigm, which allows {\balsaimpl} to quickly recognize the characteristics of poor-performing plans and adapt accordingly in future episodes. This highlights the importance of training the model to identify and avoid bad plans as early as possible.

\begin{table}[h!]
\setlength{\belowcaptionskip}{-5pt} 
\caption{Number of time-outs in different situations.}
    \centering
    \small
    \begin{tabular}{cccc}
        \toprule
        Database & Environment & Balsa & {\balsaimpl} \\
        \midrule
        IMDB & \textit{Static Workload} & 54 & \textbf{5} \\
        TPC-H & \textit{Static Workload} & 127 & \textbf{0} \\
        IMDB & \textit{Workload Switch} & 1128 & \textbf{120} \\
        TPC-H & \textit{Workload Switch} & 230 & \textbf{26} \\
        IMDB & \textit{Volume Switch} & 67 & \textbf{2} \\
        TPC-H & \textit{Volume Switch} & 949 & \textbf{0} \\
        IMDB & \textit{Both Switch} & 605 & \textbf{120} \\
        TPC-H & \textit{Both Switch} & 322 & \textbf{98} \\
        \bottomrule
    \end{tabular}
    
    \label{tab:timeout_number}
    \vspace{-5pt}
\end{table}

\subsubsection{Extra Training Overhead}
As discussed in Section~\ref{sec:q_training}, \oursystem introduces additional training overhead due to experience buffer replay and episodic updates. Although this adds cost per iteration, the overall training time remains competitive due to faster convergence. With access to high-performance GPUs such as the NVIDIA 4090 or A100, this additional overhead is further mitigated and does not hinder practical deployment. In our IMDB experiments, episodic training in \balsaimpl takes approximately 3 seconds per episode, with up to 10 episodes per iteration. The post-iteration training phase in both Balsa and \balsaimpl (including buffer replay) ranges between 4 and 9 seconds per iteration, depending on the workload (e.g., JOB, CEB, BaoQs). While \balsaimpl incurs slightly higher per-iteration training costs, it converges significantly faster, typically within 30 iterations (as shown in Figure~\ref{fig:derivative_graph})—resulting in a total training overhead of 160–310 seconds. In contrast, Balsa often requires at least 50 iterations, leading to a total overhead of 200–450 seconds. Thus, despite the added per-iteration cost, \balsaimpl achieves lower overall training time. For simpler workloads like TPC-H, the training overhead becomes minimal for both systems.

\vspace{-4pt}
\section{Related Work}
\label{section:Related Work}

\textbf{Learned Query Optimizers (LQOs).} Existing LQOs like Bao~\cite{marcus2021bao}, LEON~\cite{chen2023leon}, and Lero~\cite{zhu2023lero}, leverage neural networks to refine and enhance query plans generated by traditional optimizers. It minimally interferes with the underlying optimizer logic, reducing the risk of generating bad plans. However, these LQOs remain capped with the recommendations of traditional optimizers. On the other hand, LQOs like Balsa~\cite{yang2022balsa}, Neo~\cite{marcus2019neo}, and Lemo~\cite{mo2023lemo} control the plan generation process, allowing them to explore a broader range of execution strategies. While this leads to the discovery of superior plans, it also introduces greater variability and potential instability in performance. \oursystem can be integrated with both categories of LQOs and allow their {\lcp}s to operate in a lifelong setup, improving their training stability, and ensuring more consistent performance across varying workloads and data distributions.

\noindent\textbf{Query Execution in Dynamic Environments.} Bao~\cite{marcus2021bao} and HybridQO~\cite{yu2022hybridqo} evaluate their performance under dynamic workloads, data, and schema by periodically introducing new query templates, data updates, and schema normalization. Similarly, Lero~\cite{zhu2023lero}, LEON~\cite{chen2023leon}, and ERASER~\cite{weng2024eraser} test their models by incrementally adding new data to the database. However, they provide limited insights into the {\lqo}s' performance stability, as the updates are neither extensive nor frequent enough to fully capture the complexities of real-world, fluctuating environments. In contrast, \oursystem introduces a comprehensive benchmarking framework for dynamic scenarios, simulating more intricate changes and showing how it can enhance performance stability. Meanwhile, recent research in learned cardinality estimation has addressed challenges related to workload drifts during query execution by partially masking query information to promote generalization~\cite{negi2023robust}, utilizing rapid re-training with replay buffers~\cite{robustce24}, and applying attention mechanisms to capture relationships between queries and dynamic underlying data~\cite{li2023alece}. Conversely, \oursystem takes a more holistic approach, focusing on learned cost prediction for entire query plans.

\section{Conclusion and Future Work}
\label{section:Future Work}

In this paper, we introduce {\oursystem}, the first modular lifelong learning framework for query optimization designed to handle significant and frequent shifts in workloads and data distributions. {\oursystem} provides a lifelong learned plan cost predictor that can adapt over time while retaining and leveraging previously acquired knowledge to ensure stable performance. {\oursystem} has several contributions, including a modular plan decomposition, an attention-based neural composition, and an efficient two-phase training approach. We integrated {\oursystem} with two LQOs, and evaluated them with IMDB and TPC-H queries under dynamic situations. Our evaluation shows that \oursystem can improve query speed up to $40\%$ and the variance in running times of the same query by $60\%$ for IMDB. For TPC-H, \oursystem achieves more than $2\times$ speedup and $100\times$ query stability gain. For future work, we plan to 1) integrate {\oursystem} with additional {\lqo}s, 2) automatically identify workload-customized break operators for query decomposition and sizes for module hubs, and 3) explore knowledge transfer across different databases. In this study, we only focused on query optimization, a task with inherent modularity that aligns well with the principles of lifelong modular learning. However, we believe that our proposed framework can be extended to other learned database tasks, such as knob tuning~\cite{db-bert,lao2023gptuner} and learned indexes~\cite{ding2020alex,kraska2018case}, where modular representations and continual adaptation are also essential.

\balance
\newpage
\bibliographystyle{ACM-Reference-Format}
\bibliography{references}
\newpage

\end{document}